\documentclass[epjCONF]{svjour}
\usepackage{graphicx}
\usepackage[varg]{txfonts} % Times fonts
\usepackage[latin1]{inputenc}
\session-title{%
19$^{\textnormal{\footnotesize th}}$ International %
IUPAP Conference on Few-Body Problems in Physics%
}
\def\he3{$^3$He}
\def\al{\alpha}
\def\be{\beta}
\def\ga{\gamma}
\def\De{\Delta}
\def\de{\delta}
\def\si{\sigma}
\def\W{\Omega}
\def\w{\omega}
\def\higs{HI$\vec{\gamma}$S}
\def\bra{\langle}
\def\ket{\rangle}
\def\cpt{$\chi$PT }

\def\an{\alpha^{(n)}}
\def\bn{\beta^{(n)}}

\newcommand{\veps}{{\hat\epsilon}}
\newcommand{\vepsprime}{{\hat\epsilon\, '}}

\def\3d{3-D}

\newcommand{\dd}{\ensuremath{\mathrm{d}}}
\newcommand{\hq}{\hspace*{0.5em}}

\begin{document}
\title{Compton Scattering on Light Nuclei
}%
\author{%
% add email of the responsible author:
Deepshikha Shukla\inst{1}\fnmsep\thanks{\email{dshukla@physics.unc.edu}}
}
\institute{%
The University of North Carolina, Chapel Hill, North Carolina, USA.
}
\abstract{
Compton scattering on light nuclei ($A=2,3$) has emerged as an effective avenue to search for signatures of
neutron polarizabilities, both spin--independent and spin--dependent ones. In this discussion I will focus on
the theoretical aspect of Compton scattering on light nuclei; giving first a brief overview and therafter
concentrating on our Compton scattering calculations based on Chiral effective theory at energies of the order
of pion mass. These elastic $\gamma$d and $\gamma$He-3 calculations include nucleons, pions as the basic
degrees of freedom. I will also discuss $\gamma$d results where the $\Delta$-isobar has been included
explicitly. Our results on unpolarized and polarization observables suggest that a combination of experiments
and further theoretical efforts will provide an extraction of the neutron polarizabilities.
} %end of abstract
\maketitle
%
%
%----- Beginning of MAIN TEXT  ---------------------------------------
%
\section{Introduction}
\label{sec:intro}

Neutron structure is governed by strong-interation dynamics and hence its
electromagnetic properties encode information that contribute to our understanding
of Quantum ChromoDynamics (QCD). For example, an early
success of the $SU(3)$ quark picture was its prediction of magnetic
moments, $\vec{\mu}$, for the neutron and other strongly-interacting
particles (hadrons). Magnetic moments are a first-order response to
an applied magnetic field. In this discussion however, the focus is on
{\it electromagnetic polarizabilities}. The
two most basic polarizabilities are the electric and magnetic ones,
$\alpha_{E1}$ and $\beta_{M1}$, which quantify the second-order response of an object to
electric or magnetic field (that produces an induced dipole moment). The
Hamiltonian for a neutral particle (in this case, a neutron) in applied electric and magnetic
fields, $\vec{E}$ and $\vec{B}$, is then:
\begin{equation}
H=- \vec{\mu}\cdot \vec{B}-2\pi\left[{\alpha_{E1}(\w)}\,\vec{E}^2+
{\beta_{M2}(\w)}\,\vec{B}^2\right]. \label{eq:H1}
\end{equation}
Eq.~(\ref{eq:H1}) contains terms that are quadratic in the applied electromagnetic field.
If we now consider the derivatives (first-order) of the applied fields, four new structures appear which are second order in
$\vec{E}$ and $\vec{B}$~\cite{barry} and are explicitly dependent on the intrinsic spin of the object. They are:
\begin{eqnarray}
&&-2\pi
\left[\gamma_{E1E1}\,\vec{\sigma}\cdot\vec{E}\times\dot{\vec{E}} +
 \gamma_{M1M1}\,\vec{\sigma}\cdot\vec{B}\times\dot{\vec{B}}\right. \nonumber\\
&& - \left. 2\,\gamma_{M1E2}\,\sigma_i\,E_{ij}\,B_j  +
2\,\gamma_{E1M2}\,\sigma_i\,B_{ij}\,E_j \right] \label{eq:H2}
\end{eqnarray}
with ${T_{ij}:=\frac{1}{2} (\de_iT_j + \de_jT_i)}$. The coefficients
$\gamma_{\ldots}$s are the so-called ``spin polarizabilities''. Eqs.
(\ref{eq:H1}) and (\ref{eq:H2}) encode the multipole
parameterizations of the polarizabilities and any other
representation of the polarizabilities (for instance, spin
polarizabilities $\gamma_1$--$\gamma_4$) can be expressed as linear
combinations of these polarizabilities. This discussion intends to
show that for the neutron, these six polarizabilities can be
extracted from Compton scattering on on the deuteron and \he3.

Polarizabilities such as those in Eqs.~(\ref{eq:H1}) and
(\ref{eq:H2}) can be accessed in Compton scattering because
the effective Hamiltonian yields a nucleon Compton scattering amplitude of the form:
\begin{equation}
T_{\ga N}= \sum \limits_{i=1 \ldots 6} A_i (\w,\, \theta) t_i.
 \label{eq:amp}
\end{equation}
Here $t_1$--$t_6$~\cite{bkmrev} are invariants constructed out of the photon
momenta and polarization vectors ($\veps$ and $\vepsprime$). $t_1$ and $t_2$ contain terms that are
nucleon-spin independent whereas $t_3$--$t_6$ include nucleon spin. The $A_i$'s are Compton structure functions;
expanding these $A_i$ around $\w=0$ illustrates the how the polarizabilities affect the Compton amplitude. $\alpha$
$\beta$ enter at $\mathcal{O}(\w^2)$ whereas the spin-polarizabilities enter at $\mathcal{O}(\w^3)$.
For the proton, the Thomson term, $-\frac{Q^3}{M} \vepsprime \cdot
\veps$ ensures a larger cross section (compared to the neutron) and from  low-energy Compton scattering measurements
$\alpha^{(p)}$ and $\beta^{(p)}$ can be extracted. A combined analysis of the differential cross-sections (dcs) of a
number of $\ga$p experiments over the past decade~\cite{Schumacher06} yields the PDG values:
\begin{eqnarray}
\alpha_p&=&(12.0 \pm 0.6) \times 10^{-4} \, {\rm fm}^3, \nonumber \\
\beta_p&=& (1.9 \pm 0.5) \times 10^{-4} \, {\rm fm}^3. \label{eq:pexp}
\end{eqnarray}

Similar extractions for the neutron is not possible because
\begin{itemize}
 \item the neutron Thomson term is absent and,
 \item there are no free-neutron targets as neutrons are very short-lived (lifetime $\sim$880 secs.).
\end{itemize}
Of all the possible ways to access neutron polarizabilities (that includes scattering neutrons off a heavy nucleus
such as lead), Compton scattering on light nuclei has emerged as likely candidates that would enable the extraction
of the six neutron polarizabilities. For instance, the latest global analysis of the $28$ points for deuteron Compton
scattering gave
\begin{eqnarray}
  \label{eq:neutronpols}
  {\alpha}^s&=&11.3\pm0.7_\mathrm{stat}\pm0.6_\mathrm{Baldin}\pm1_\mathrm{th}
  \nonumber \\
  {\beta}^s &=&3.2\mp0.7_\mathrm{stat}\pm0.6_\mathrm{Baldin}\pm1_\mathrm{th}
\end{eqnarray}
for the iso-scalar nucleon polarizabilities~\cite{Hi05b,Hi05} with the Baldin
sum rule $ \bar{\alpha}^{(s)}+\bar{\beta}^{(s)}=14.5\pm0.6$ as constraint.
The situation for the neutron spin polarizabilities is much worse
and the only data available are for the forward and the backward
spin polarizabilities, which are linear combinations of the four
spin polarizabilities. The neutron backward spin polarizability was
determined to be
\begin{eqnarray}
\ga_{\pi} &=& -\ga_{E1E1}-\ga_{E1M2}+\ga_{M1E2}+\ga_{M1M1} \nonumber \\
&=& (58.6 \pm 4.0) \times 10^{-4} \, {\rm fm}^4, \label{eq:gpn}
\end{eqnarray}
from quasi-free Compton scattering on the deuteron~\cite{Ko03}. The
\cpt prediction for $\ga_{\pi}$ is $57.4 \times 10^{-4}$
fm$^4$~\cite{Pa03}.
The forward spin polarizability, $\ga_0$, is related to
energy-weighted integrals of the difference in the
helicity-dependent photoreaction cross-sections ($\si_{1/2} -
\si_{3/2}$). Using the optical theorem one can derive the following
sum rule~\cite{bkmrev,ggt}:
\begin{eqnarray}
\ga_0 &=& -\ga_{E1E1}-\ga_{E1M2}-\ga_{M1E2}-\ga_{M1M1} \nonumber \\
&=&  \frac{1}{4
\pi^2} \int_{\w_{th}}^{\infty} \frac{\si_{1/2}-\si_{3/2}}{\w^3}\,
\mathrm{d}\w, \label{eq:g0}
\end{eqnarray}
where $\w_{th}$ is the pion-production threshold. The following
results on $\ga_0$ were obtained using the VPI-FA93 multipole
analysis~\cite{Sa94} to calculate the integral on the RHS of
Eq.~(\ref{eq:g0}):
\begin{equation}
\ga_{0} \simeq -0.38 \times 10^{-4} \, {\rm fm}^4. \label{eq:g0n}
\end{equation}
The \cpt prediction for $\ga_{0}$ is consistent with
zero~\cite{Pa03}.

New data for unpolarized deuteron Compton scattering from MAXlab is being analysed~\cite{myers}, and
an experiment at HI$\gamma$S is approved.
Additionally, the fact that a polarized \he3~nucleus behaves as an effective neutron~\cite{he3pol} has generated
interest in \he3~Compton scattering. The first elastic $\gamma$\he3~calculations were reported in Ref.~\cite{Ch06,Ch07,Sh09}
and preparations are underway for a proposed experiment at \higs. Currently, a concerted effort is 
underway to reduce the theory-error using
higher orders in the chiral counting~\cite{allofus}. The goal is a
comprehensive approach to Compton scattering in the proton~\cite{Be02,Hi04},
deuteron~\cite{Hi05b,Hi05,Be02,Be99,Hi05a,Ch05} and ${}^3$He~\cite{Ch06,Ch07,Sh09} in
$\chi$EFT from zero energy to beyond the pion-production threshold.
These proceedings give a quick overview of the current investigations
to help extract the neutron polarizabilities from polarized and unpolarized deuteron and \he3~Compton
scattering.

\section{Compton scattering off a nucleon}
\label{sec:gammaN}

\begin{figure}[!htb]
\begin{center}
  \includegraphics*[width=0.95\linewidth]{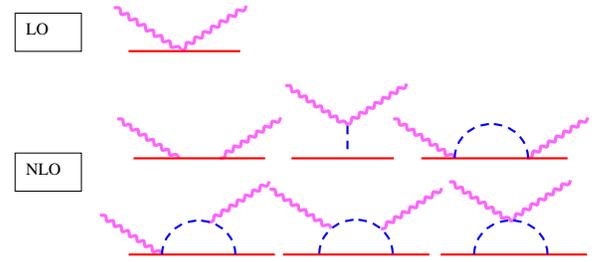}
  \caption{\label{fig:oq3N}The LO and NLO contributions to nucleon Compton scattering. The
  wiggly lines are photons, dashed ones are pions and the solid ones are nucleons. Permutations and
    crossed diagrams not shown.}
\end{center}
\end{figure}

In order to calculate the Compton amplitude (\ref{eq:amp}) at energies of the order of the pion mass, we employ
Heavy Baryon Chiral Perturbation Theory (HB\cpt) with pions and nucleons as
effective degrees of freedom~\cite{bkmrev}. The expansion parameter in this formulation is 
$Q=\left(\frac{p,m_{\pi}}{M,\Lambda_{\chi}}\right)$, where $p$ is a small momentum usually comparable
to the pion mass, $m_{\pi}$; $M$ is the nucleon mass and $\Lambda_{\chi}$ is the scale of chiral symmetry 
breaking. Fig.~\ref{fig:oq3N}
shows the leading and the next-to-leading order contributions in this framework. At
this order the theory is parameter-free and the polarizabilities are
then predictions of the theory.
The nucleon Compton amplitude at NNLO has also been
calculated~\cite{Mc00}, however for the purpose of the results
reported in these proceedings they are immaterial.

\begin{figure}[!htb]
\begin{center}
  \includegraphics*[width=0.95\linewidth]{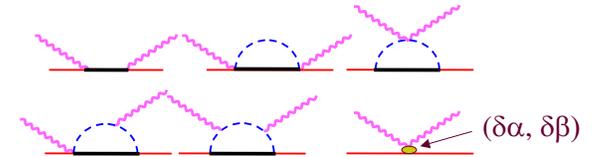}
  \caption{\label{fig:deltagraphs}The $\De$ contributions to nucleon Compton scattering up to NLO. The
  thick solid ones are depict the intermediate $\De$-isobar.}
\end{center}
\end{figure}
The $\Delta$-isobar can be included explicitly in the calculation
and the results reported in these proceedings employ the ``Small
Scale Expansion" (SSE) scheme~\cite{He97}. In this scheme the expansion parameter, $\varepsilon$ denotes either a 
small momentum, the pion mass, or the mass difference $\Delta_0$ between the real
part of the $\Delta$ mass and the nucleon mass, $i.e.\,
\Delta_0={\mathrm{Re}}(M_{\Delta}) -M$.
Fig. \ref{fig:deltagraphs} shows the
$\De$-contributions up to the NLO. At this order, the
spin-polarizabilities are still parameter-free predictions. However,
the large paramagnetic $\De$-effects and the included pion
contributions are not sufficient to explain the observed proton
magnetic polarizability. This necessitates the introduction of two
low-energy coefficients $\delta\alpha,\;\delta\beta$. These
``off-sets'' are then determined from data. It should be mentioned here that the $\De$-isobar 
has been included in the $\ga$d calculation only.

\section{The nucleon inside a nucleus}
\label{sec:gammaA}

The $\ga$-nucleus scattering amplitude is written as
\begin{equation}
{\mathcal M}=\bra \Psi_f|{\hat O}|\Psi_i \ket , \label{calM}
\end{equation}
with $|\Psi_i\ket$ and $|\Psi_f\ket$ being the nuclear
wavefunctions. ${\hat O}$ represents the photon-nucleon(s)
interaction kernel that is calculated using \cpt. The nuclear
wavefunctions are usually derived from a potential model or a chiral
potential.

\begin{figure}[!htbp]
  \begin{center}
    \includegraphics*[width=\linewidth]{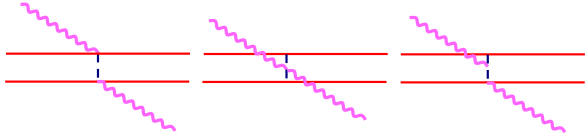}
    \caption{\label{fig:2bnlographs} Lowest-order two-body current contributions. They enter a $\ga$-nucleus
    calculation at NLO. Crossed graphs and graphs with the nucleons interchanged are not shown.}
\end{center}
\end{figure}

Neutron properties are usually extracted from data taken on few-nucleon
systems by dis-entangling nuclear-binding effects. Within the
power-counting scheme used for these calculations, these effects
include pion-exchange mechanisms between two nucleons at the lowest
order (see Fig.~\ref{fig:2bnlographs}). The external photon may
either couple to a pion-nucleon vertex or an exchanged pion. The
graphs shown in Fig.~\ref{fig:2bnlographs} enter the $\ga$d or
$\ga$\he3~calculation only at next-to-leading order (${\mathcal O} (Q^3)$).
Additional contributions at NNLO have also been
calculated~\cite{Be02} but are not essential for this discussion.

\section{Observables}

\begin{figure}[!htb]
\begin{center}
\small{{linpol.~$\gamma$, unpol.~tar.:}}
{$\displaystyle\left.\frac{\dd\sigma}{\dd\Omega}\right|^\mathrm{lin}_x$} \hq
\includegraphics*[width=0.2\linewidth]{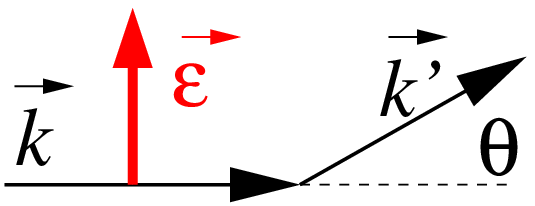}
\hq\hq,\\
{$\displaystyle\left.\frac{\dd\sigma}{\dd\Omega}\right|^\mathrm{lin}_y$} \hq
\includegraphics*[width=0.2\linewidth]{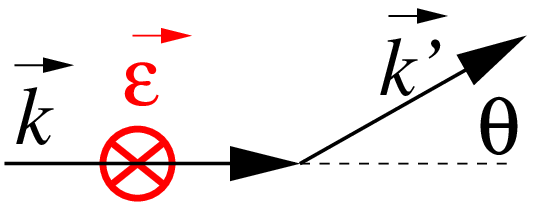}\\[2ex]
\parbox{10ex}{{circpol.~$\gamma$,\\ vecpol.~tar.}}:
{$\displaystyle\Delta_x^\mathrm{circ}$}
\hq\parbox{0.7\linewidth}{
\includegraphics*[width=\linewidth]{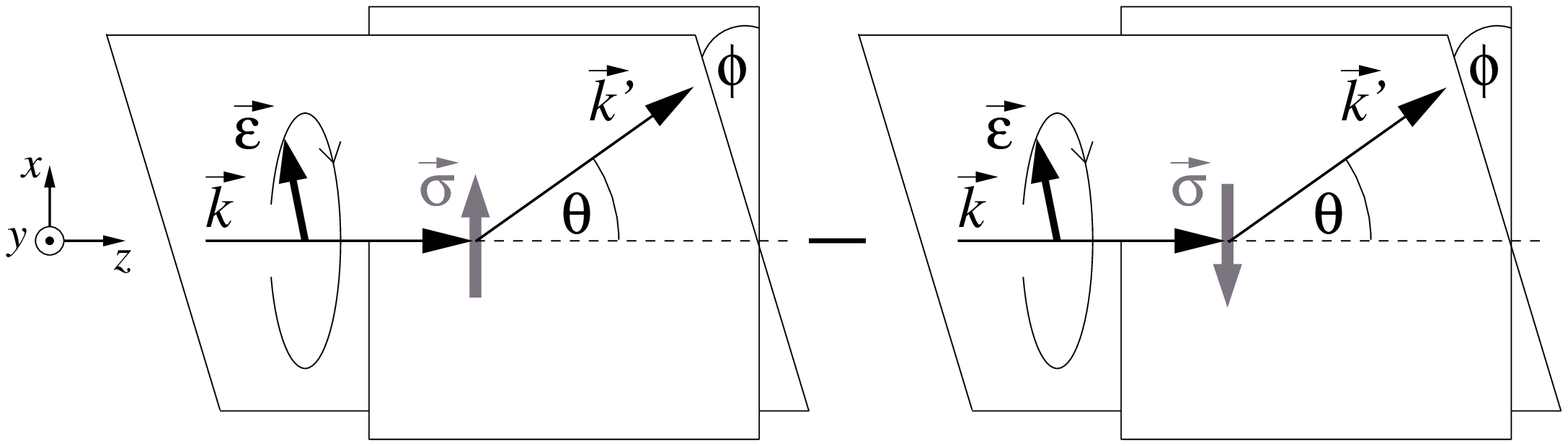}}
\hq\hq,\\
{$\displaystyle\Delta_z^\mathrm{circ}$}
\hq\parbox{0.7\linewidth}{
\includegraphics*[width=\linewidth]{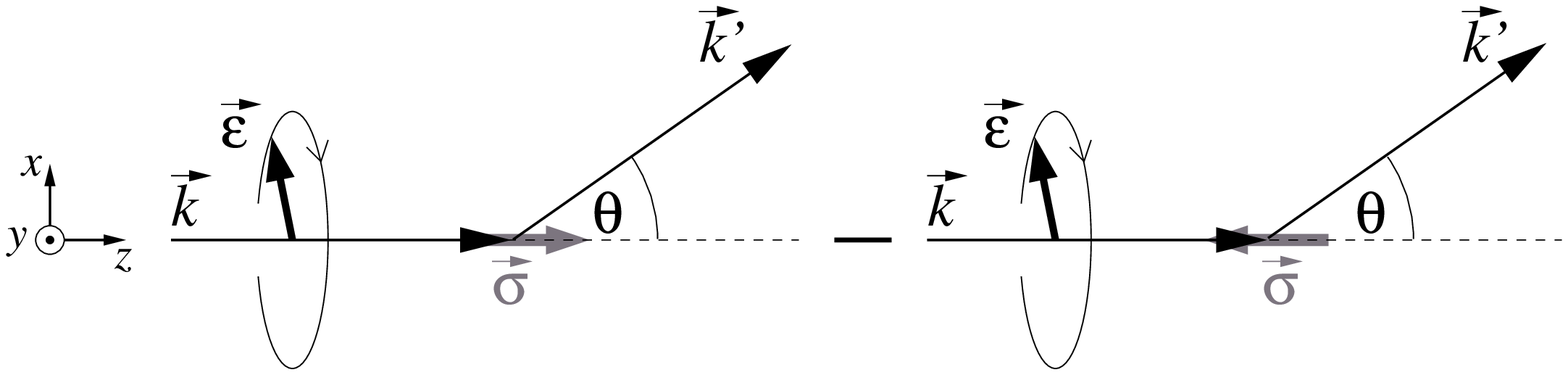}}\\[2ex]
\parbox{10ex}{{linpol.~$\gamma$,\\ vecpol.~tar.}}:
{$\displaystyle\Delta_x^\mathrm{lin}% ,\;
% \Sigma_x^\mathrm{circ}=\frac{\Delta_x^\mathrm{circ}^\mathrm{lin}}{\mathrm{sum}}
$}
\hq\parbox{0.7\linewidth}{
\includegraphics*[width=\linewidth]{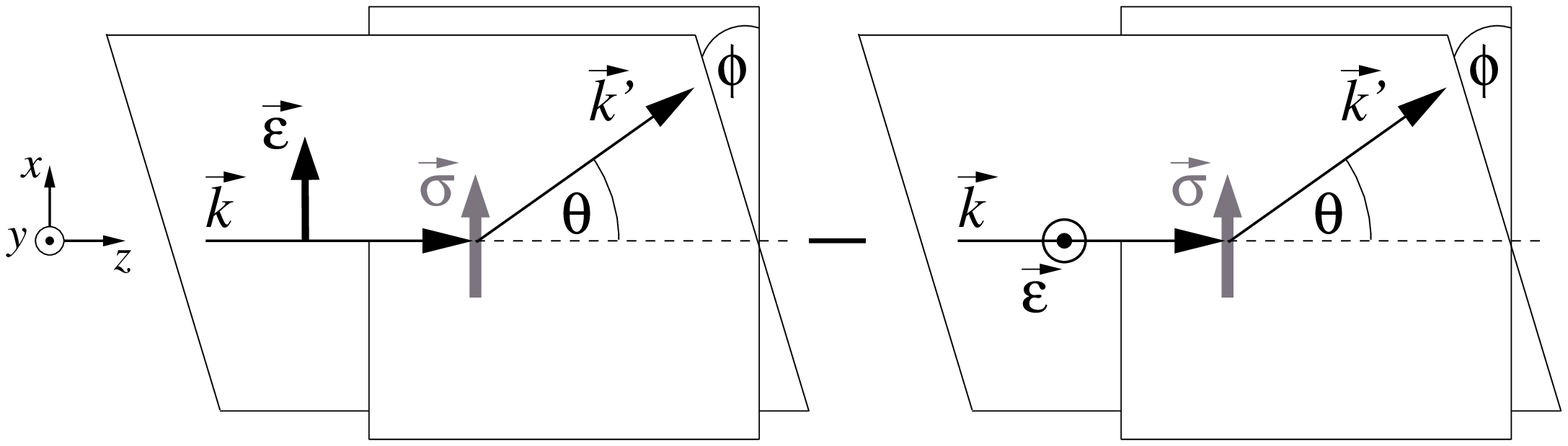}}
\hq\hq,\\
{$\displaystyle\Delta_z^\mathrm{lin}% ,\;
% \Sigma_z^\mathrm{circ}=\frac{\Delta_z^\mathrm{circ}^\mathrm{lin}}{\mathrm{sum}}
$}
\hq\parbox{0.7\linewidth}{
\includegraphics*[width=\linewidth]{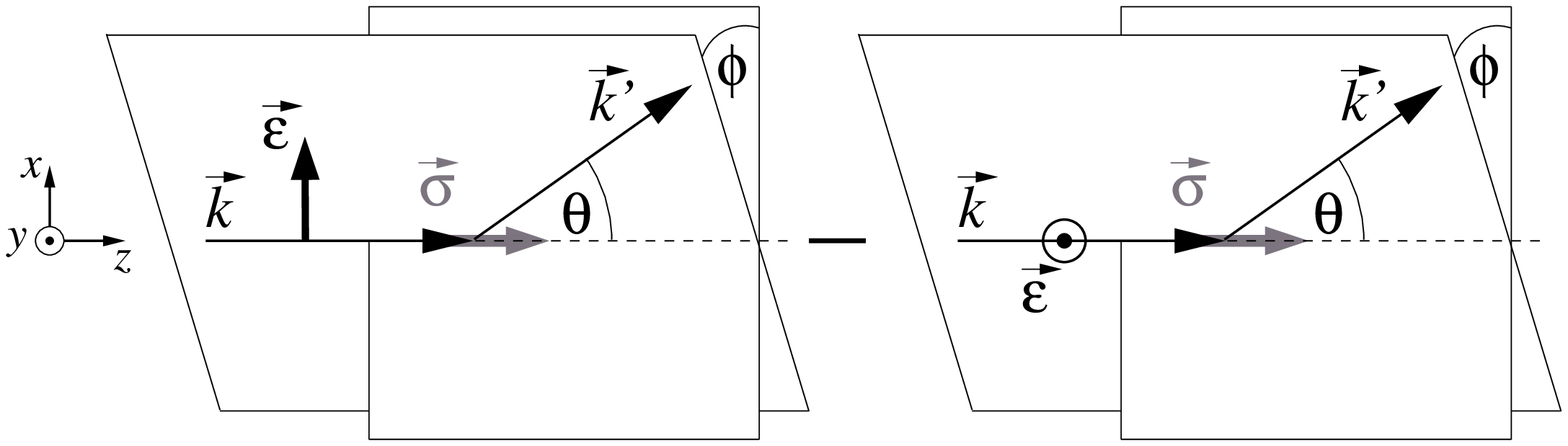}}
\caption{\label{fig:observables}Definition of observables for singly and double
  polarized observables. Figure from H.~Grie{\ss}hammer.}
\end{center}
\end{figure}

The availability of polarized beams and the technology to polarize targets has opened up the
field of possible measurements. Over and above the canonical differential cross-section, one can now 
measure observables that include different combinations of unpolarized/polarized beam and target. 
Fig.~\ref{fig:observables} gives a representation of the observables considered. The convention adopted is that
the beam direction is the $z$-axis and the $x-z$ plane is the scattering plane. For a linearly polarized beam 
and unpolarized target, $\left.\frac{\dd\sigma}{\dd\Omega}\right|_x^\mathrm{lin}$ is the differential
cross-section for photon polarization in the scattering plane, and
$\left.\frac{\dd\sigma}{\dd\Omega}\right|_y^\mathrm{lin}$ for perpendicular
polarization. The $\De$s are the double polarized observables that involve a
vector-polarized deuteron and a circularly or linearly polarized photon.
$\De_{x(z)}^{\mathrm {circ}}$ gives the difference in the difference in the differential cross-sections
between configurations when the target is polarized along $+{\hat x}$ ($+{\hat z}$) and $-{\hat x}$ ($-{\hat z}$)
using a circularly polarized beam. $\De_{x(z)}^{\mathrm {lin}}$ gives the photon polarization asymmetry when the 
target is polarized along $+{\hat x}$ ($+{\hat z}$). Experimental measurents often use the asymmetries $\Sigma$
obtained by dividing the $\De$s by the sum of the measured differential cross-sections. Of course, these asymmetries
are then devoid of systematic uncertainties.

The observables described above are only a subset of all the possible combinations. The focus of these 
proceedings is only on some prominent examples. For the deuteron, representative results for most of the above obervables
are presented. The aim is to aid in planning new
experiments, and the results for all of the above observables are available as an interactive
\emph{Mathematica 7.0} notebook from Grie{\ss}hammer (hgrie@gwu.edu). For the \he3~Compton scattering calculation
results for the unpolarized differential cross-section and $\De_{x(z)}^{\mathrm {circ}}$ only are presented.

\section{Elastic $\gamma$d scattering}
\label{sec:gammad}

The NLO (${\mathcal O}(\varepsilon^3)$) deuteron Compton scattering calculations in the SSE variant of \cpt include the 
amplitudes of Secs.~\ref{sec:gammaN} and 
\ref{sec:gammaA}. Note that the pion-pole graph of Fig.~\ref{fig:oq3N} does not contribute to 
deuteron Compton scattering as deuteron is an isoscalar. There is  an additional ingredient in the calculation -- resumming the 
intermediate $NN$-rescattering states. A consistent description of $\ga$d Compton scattering
must also give the correct Thomson limit, an exact low-energy theorem which in turn
follows from gauge invariance~\cite{Friar}. In deuteron Compton scattering, this mandates the 
inclusion of $T_{NN}$ whenever both nucleons propagate close to their mass-shell between photon absorption
and emission, i.e.~when the photon energy $\omega\lesssim 50\;\mathrm{MeV}$. At higher photon energies
$\omega\gtrsim 60\;\mathrm{MeV}$, the nucleon is kicked far enough off its
mass-shell, $E\sim Q$, for the amplitude to become perturbative. 
\begin{figure}[!htbp]
  \begin{center}
    \includegraphics*[width=\linewidth]{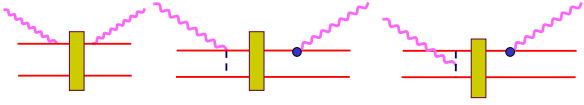}
    \caption{\label{fig:nnrescat} Low-energy $NN$ rescattering contributions to deuteron Compton scattering.
    The rectangle in the middle represents the $NN$ T-matrix.}
\end{center}
\end{figure}
Figure~\ref{fig:nnrescat} lists the additional contributions to Compton scattering off the
deuteron to next-to-leading order in $\chi$EFT when the low-energy resummation of the intermediate 
$NN$-rescattering states is performed. 

Once all the ingredients in ${\hat O}$ of Eq.~(\ref{calM}) are determined, it is folded with deuteron 
wavefunctions to obtain the deuteron Compton amplitude, using which any observable
can be calculated. For the results reported here we use chiral
NNLO wavefunctions for the deuteron and the $AV18$ potential in the
intermediate $NN$-rescattering process. While it is desirable that
for a consistent calculation the $\ga NN$ kernel, the wavefunctions
and the potential used to calculate the intermediate rescattering
contribution should be derived using the same framework. Our $\ga
NN$ kernel and the wavefunctions are indeed derived from the same
framework, but $AV18$ is not a chiral potential. However, it has
been shown in Ref.~\cite{Hi05b} and since then we have verified that the form of the potential used
in the intermediate rescattering process causes barely perceivable
difference in the final results. Issues of matching currents and
couplings etc. only appear at two higher orders than what we
calculate.

Also note that the polarizabilities extracted from deuteron Compton scattering are the isoscalar combinations. 
This means that the extraction of the neutron polarizabilities will depend on how accurately the proton polarizabilities
are known. 

\subsection{Significance of  the $\Delta$ and $NN$ Rescattering on Polarization Observables}
\label{sec:comparison}

\begin{figure*}[!htb]
\begin{center}
\includegraphics*[width=0.35\linewidth]{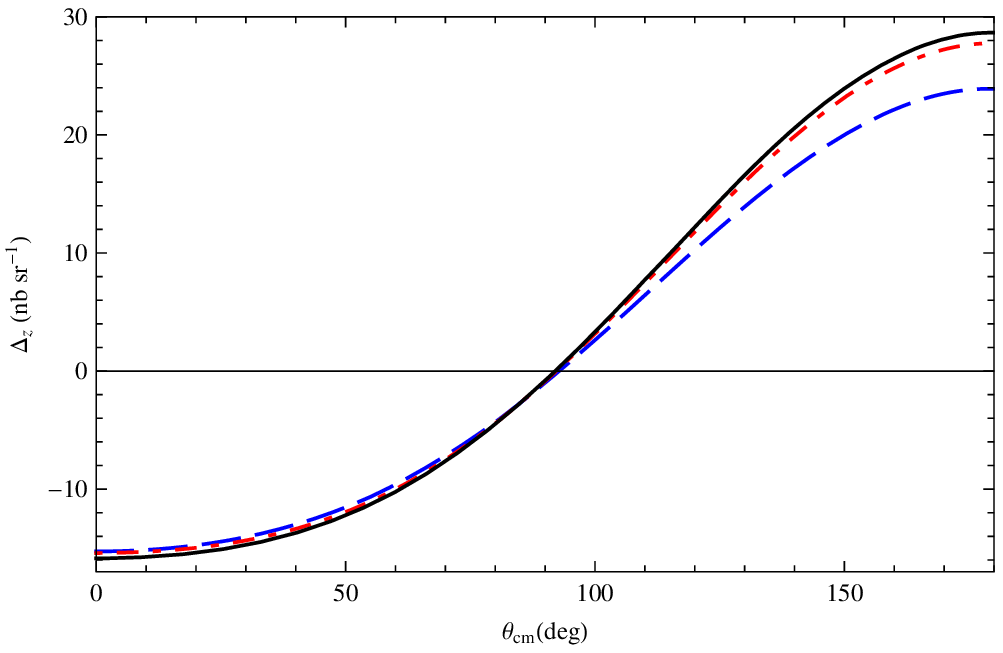}
%\hfill
\includegraphics*[width=0.35\linewidth]{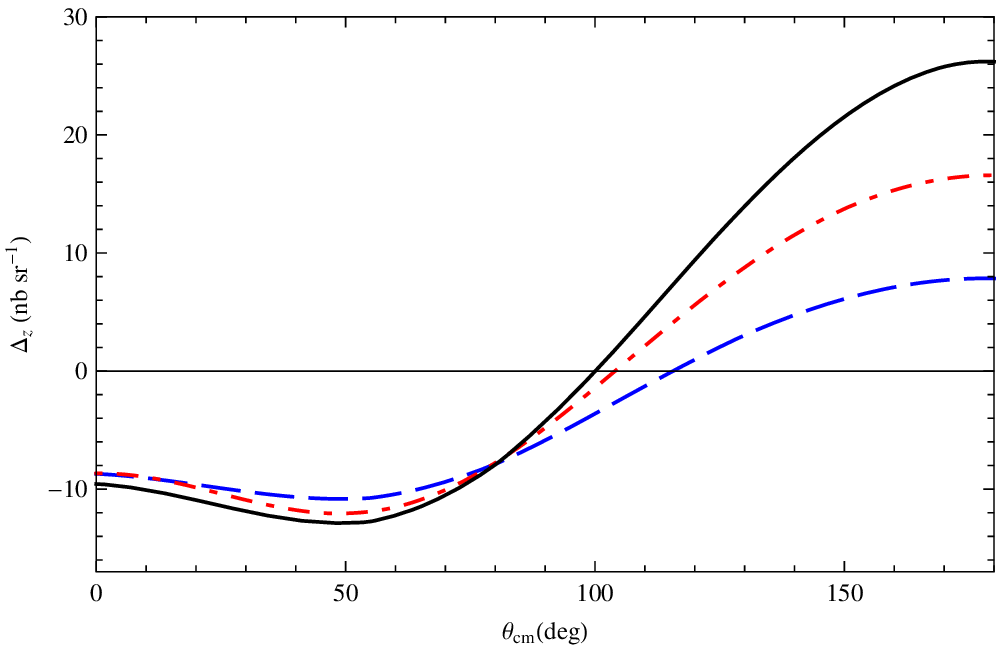}\\
\includegraphics*[width=0.35\linewidth]{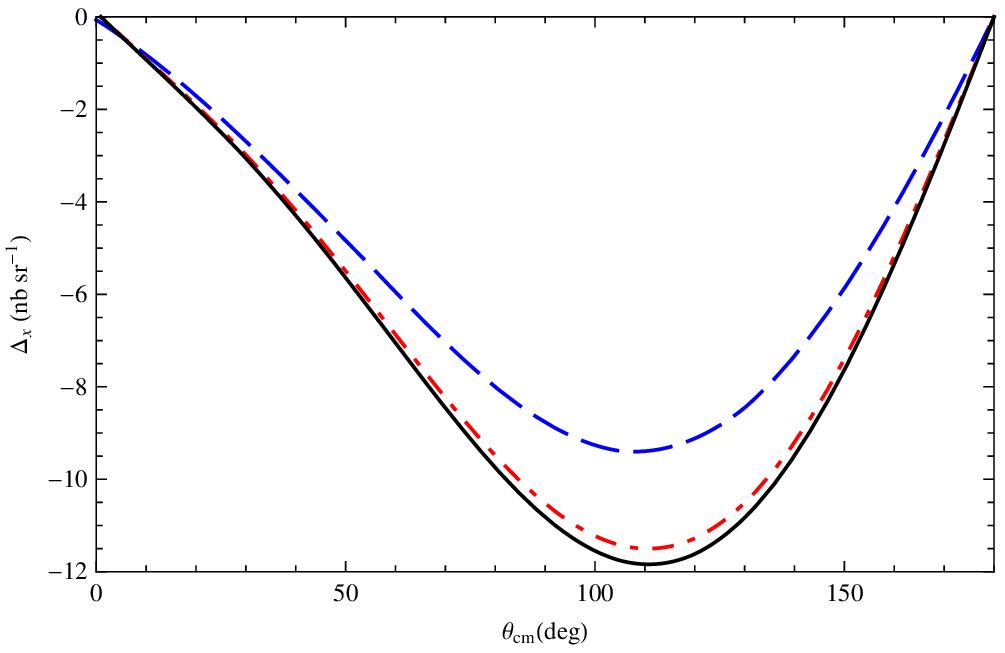}
%\hfill
\includegraphics*[width=0.35\linewidth]{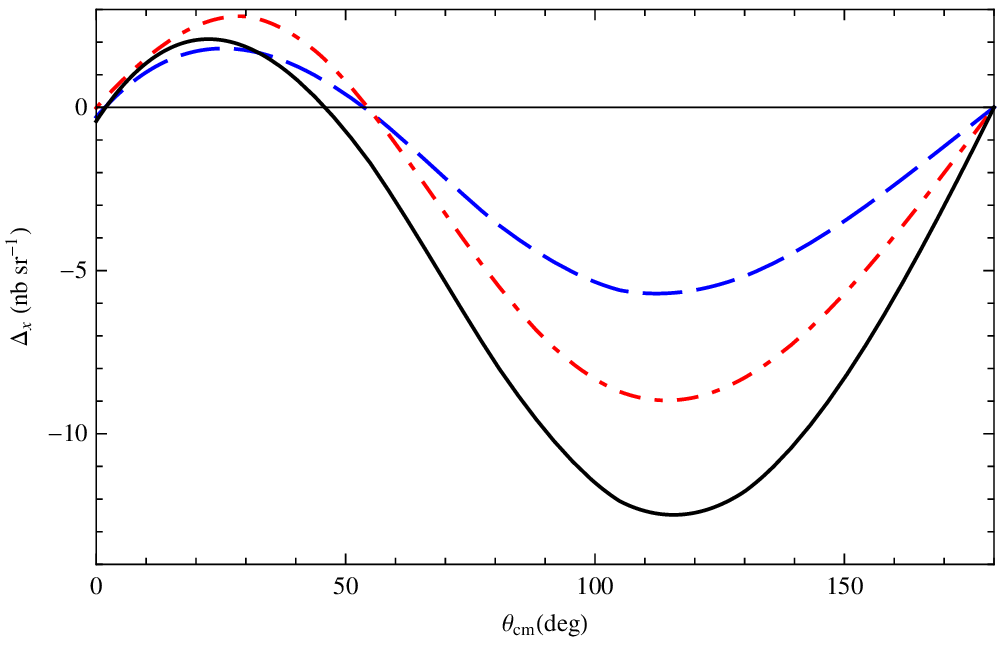}
\caption{\label{fig:comp}Effects of $\De(1232)$ and of resumming $NN$
  intermediate states on the double polarization observables $\De_z$ (top) and
  $\De_x$ (bottom). Left: $\w_\mathrm{lab}=45$ MeV; right:
  $\w_\mathrm{lab}=125$ MeV. Legend given in the text.
 2}
\end{center}
\end{figure*}

We first analyse the impact of including the $\De$-isobar and intermediate $NN$ rescattering
contributions on polarization observables.  Figure~\ref{fig:comp} compares
double-polarization observables within different schemes. The top row shows $\De_z^\mathrm{circ}$ 
(left: 45~MeV and right: 125~MeV) and the bottom row shows $\De_x^\mathrm{circ}$ 
(left: 45~MeV and right: 125~MeV). In the left-hand panels, the dashed line is a ${\mathcal O} (Q^3)$ HB\cpt
calculation without dynamical $\Delta$ or rescattering, the dot-dashed
line is the same calculation with $NN$-rescattering included, and the
solid line is a ${\mathcal O} (\varepsilon^3)$ calculation with both
intermediate $NN$ rescattering and a dynamical $\Delta$. In the right-hand panels, the dashed line is a
${\mathcal O} (Q^3)$ HB\cpt calculation without dynamical $\Delta$, the
dot-dashed line has the dynamical $\Delta$ added, but no
$NN$-rescattering, and the solid line is the full calculation with
intermediate $NN$ rescattering and a dynamical $\Delta$
As for unpolarized observables~\cite{Hi05,Hi05a}, the $\De(1232)$ does not
contribute appreciably at $45$~MeV, but the observable is still ruled by
including intermediate $NN$ rescattering for the correct Thomson limit. In
contrast, the $\De$ and intermediate $NN$-rescattering are equally
significant at $125$~MeV. % .

Thus, together with the observations of Refs.~\cite{Hi05,Hi05a} for unpolarized deuteron Compton scattering, 
we emphasize that the $\De$-isobar and the intermediate $NN$-
rescattering contributions are necessary ingredients of our
calculations that attempt to identify observables for the extraction
of neutron polarizabilities in the energy range 45 -- 125 MeV (lab). It has been verified that the dependence 
on the potential that is used to generate the deuteron wavefunctions and the $NN$ T-matrix (in the 
intermediate state) is negligible.

\subsection{Results}
\label{sec:results}

\begin{figure}[!htb]
\begin{center}
\includegraphics*[width=.85\linewidth]{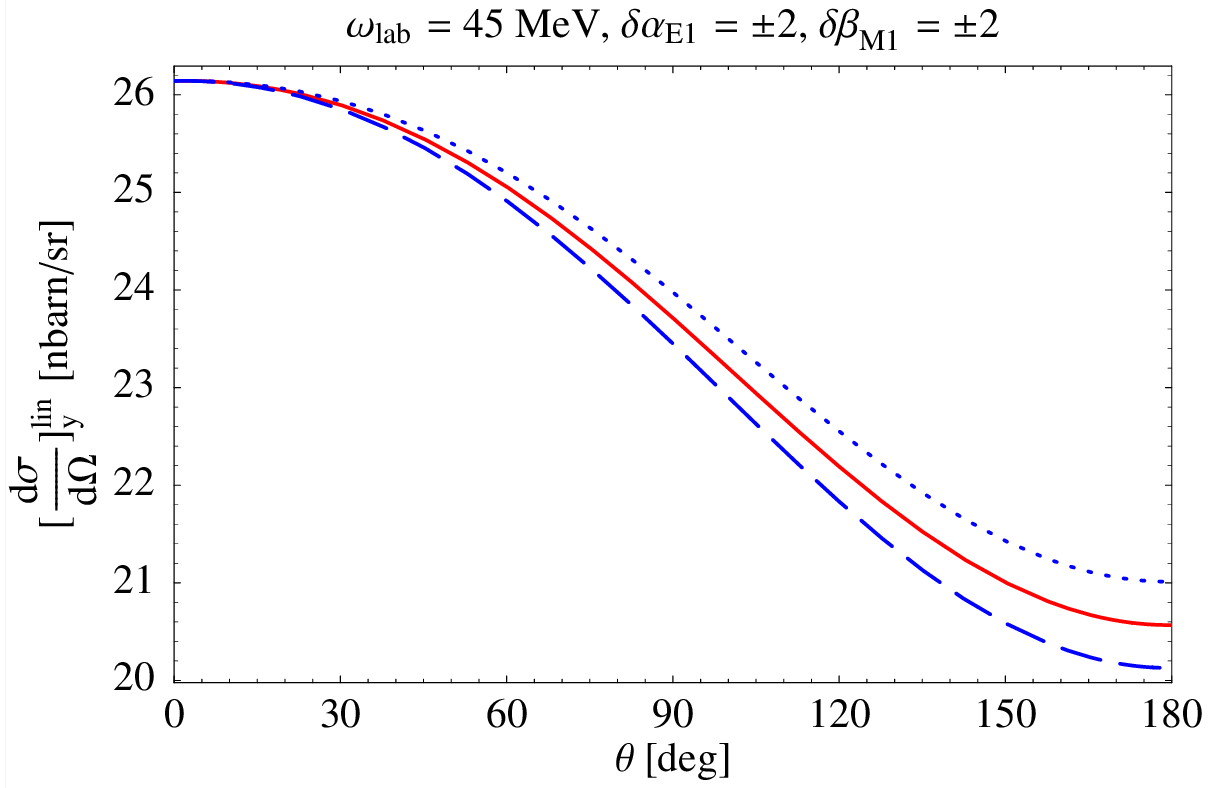}\\
\includegraphics*[width=.85\linewidth]{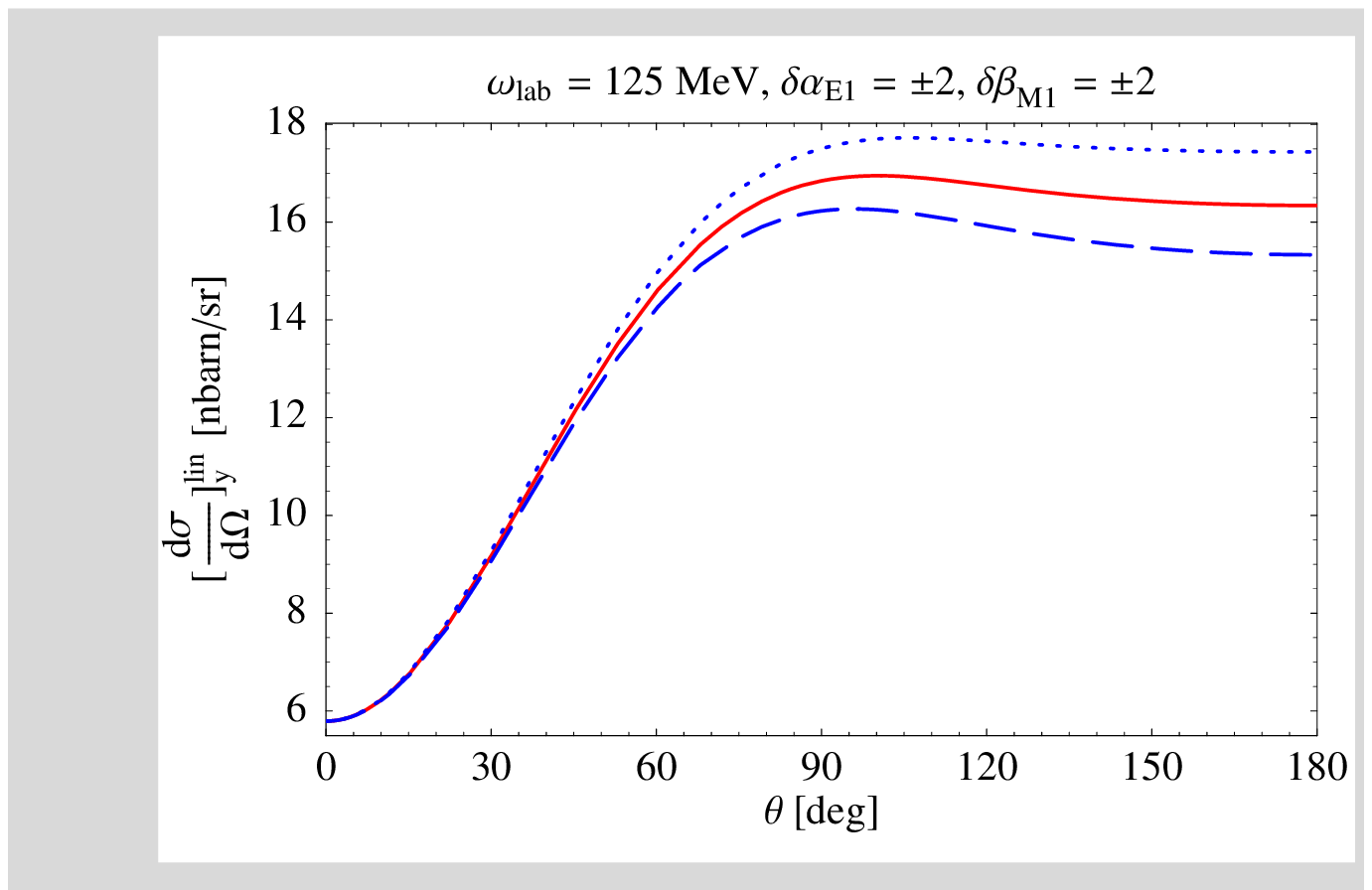}\\
\includegraphics*[width=.85\linewidth]{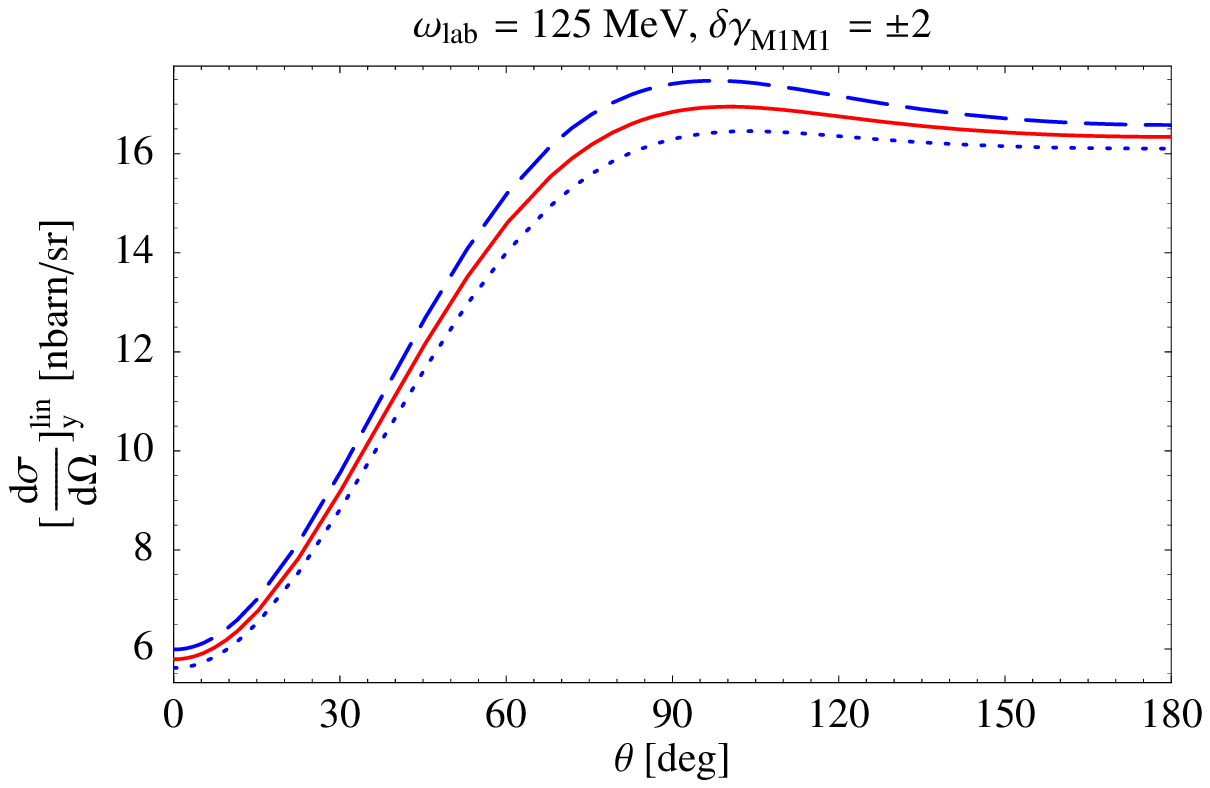}
\caption{\label{fig:dcsy} Differential cross-sections with photons
  linearly-polarized along the $y$-axis. Top and middle panels: The combination ${\alpha}^{E1}-{\beta}^{M1}$ 
  is varied, while their sum is constrained by the Baldin sum rule. Bottom panel: 
  $\ga_{M1M1}$ varied by $\pm2$ units at $\w_\mathrm{lab}=125$ MeV.}
\end{center}
\end{figure}
\begin{figure}[!htb]
\vspace{-2ex}
\begin{center}
\includegraphics*[width=.85\linewidth]{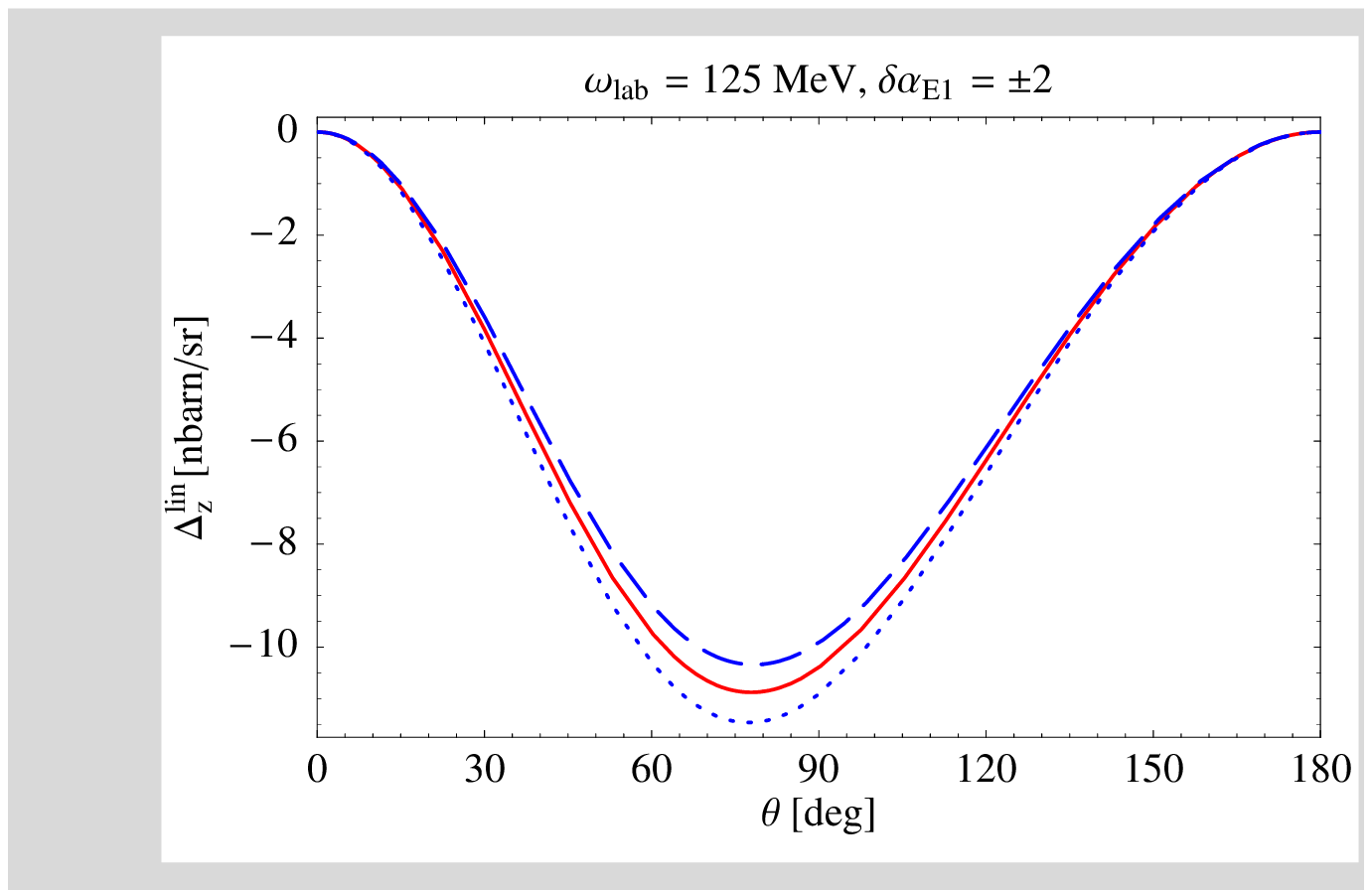}\\
\includegraphics*[width=.85\linewidth]{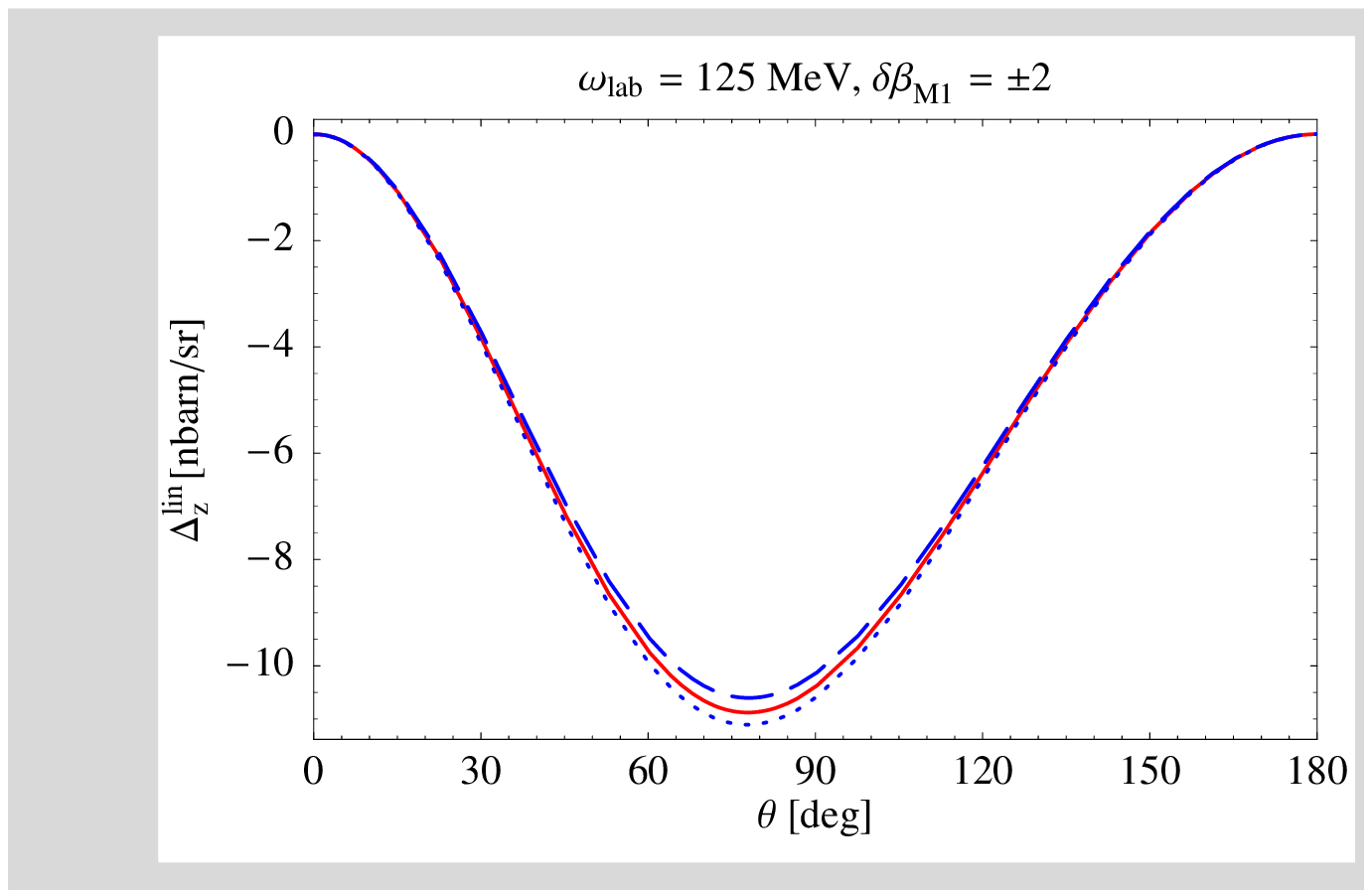}\\
\includegraphics*[width=.85\linewidth]{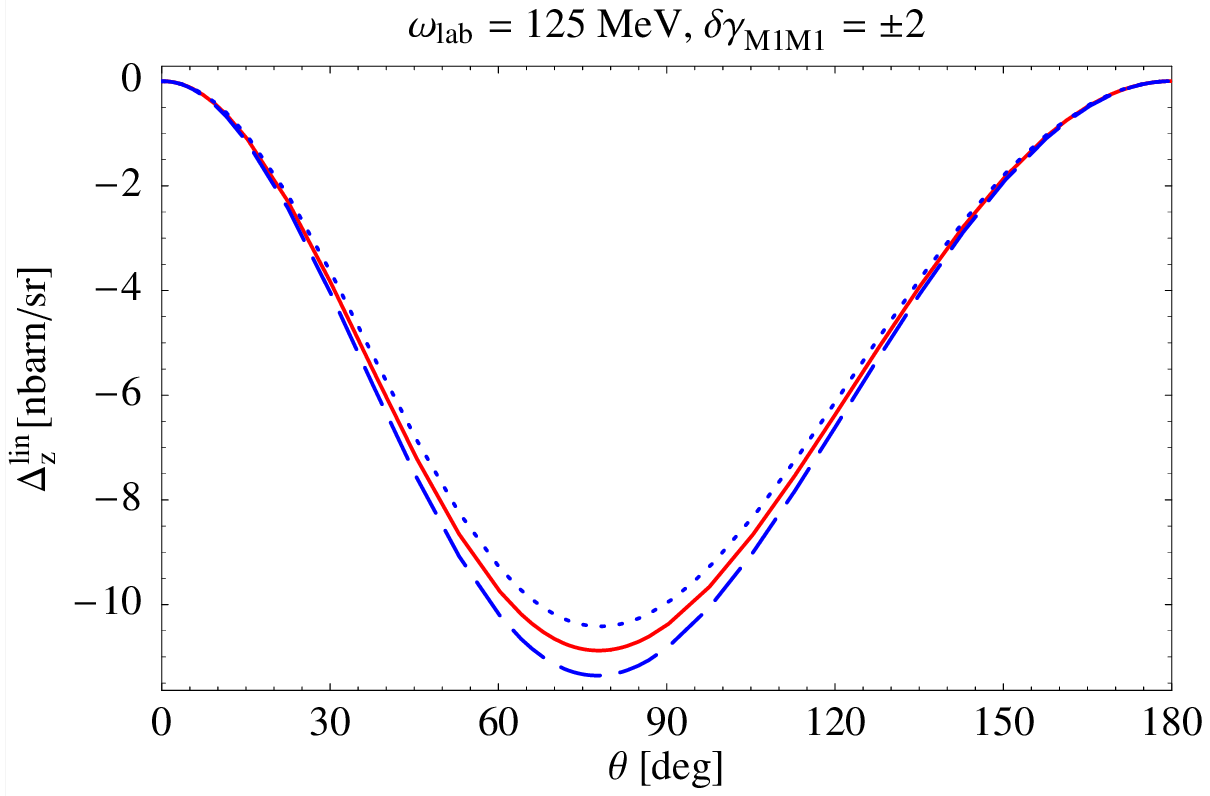}
\caption{\label{fig:deltazlin} The double-polarization observable
  $\Delta_z^\mathrm{lin}$ with linearly-polarized photons at $\w_\mathrm{lab}=125$
  MeV. ${\alpha}^{E1}$ ($\be_{M1}$) is varied by $\pm 2$ units in the topmost (middle) panel.
  The bottom panel shows variation of $\ga_{M1M1}$ by $\pm 2$.}
\end{center}
\end{figure}
We now present results for selected polarization observables which show appreciable sensitivity to the polarizabilities. 
Strictly speaking, at this order only $\al_{E1}$ and $\be_{M1}$ have an undertermined short-distance piece;
the spin-polarizailities are predictions of the theory. However, we have adopted the strategy whereby we 
arbitrarily add six parameters $\de \al_{E1}$, $\de \be_{M1}$ and $\de \ga_{ij}$s to the calculation to represent 
effects not explictly included in the theory. This 
allows us to gauge the dependence of the observables on these parameters. 

\begin{figure*}[!htb]
\vspace{-2ex}
\begin{center}
\includegraphics*[width=0.31\linewidth]{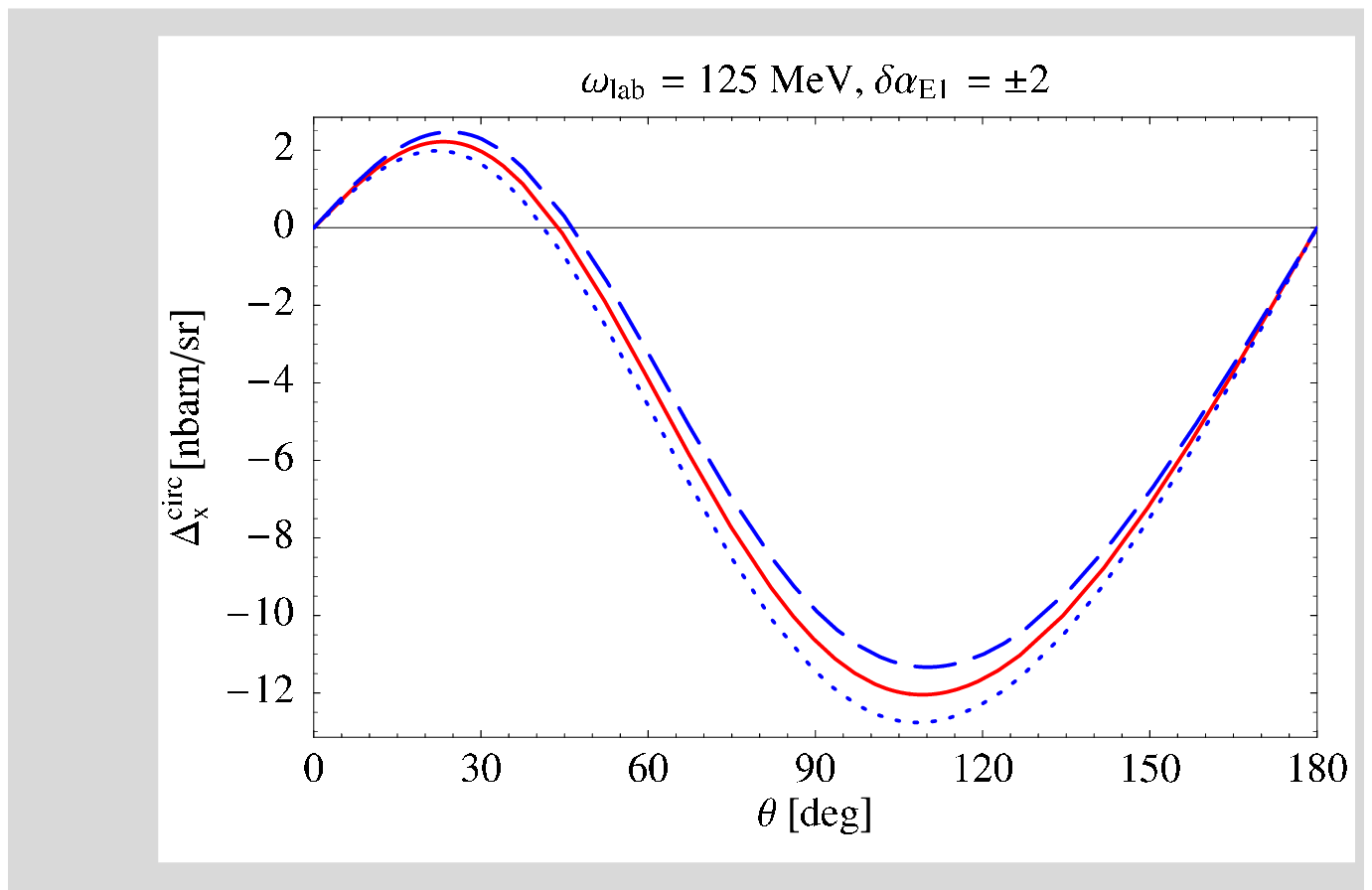}
\hq\hq
\includegraphics*[width=0.31\linewidth]{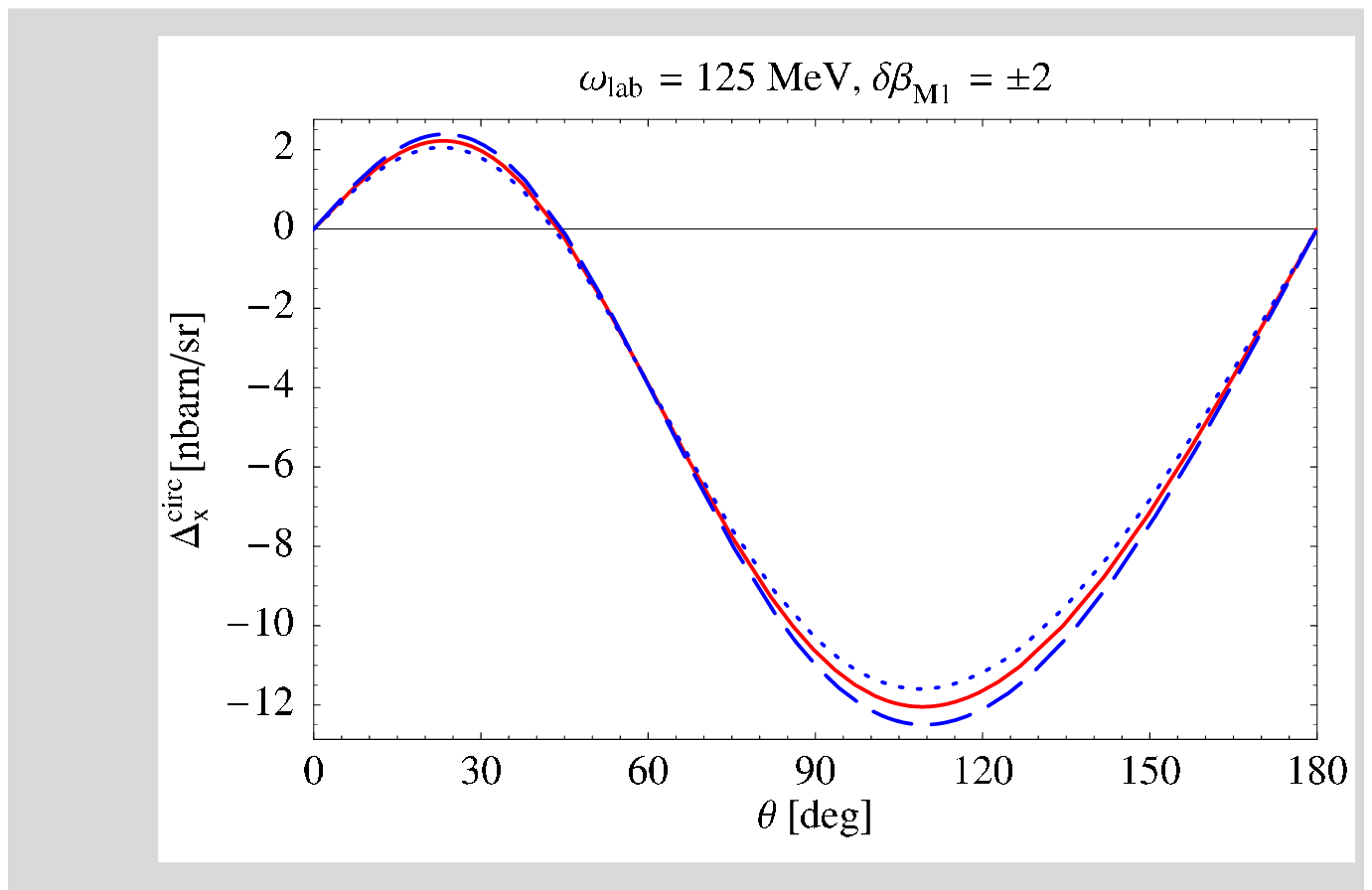}
\hq\hq
\includegraphics*[width=0.31\linewidth]{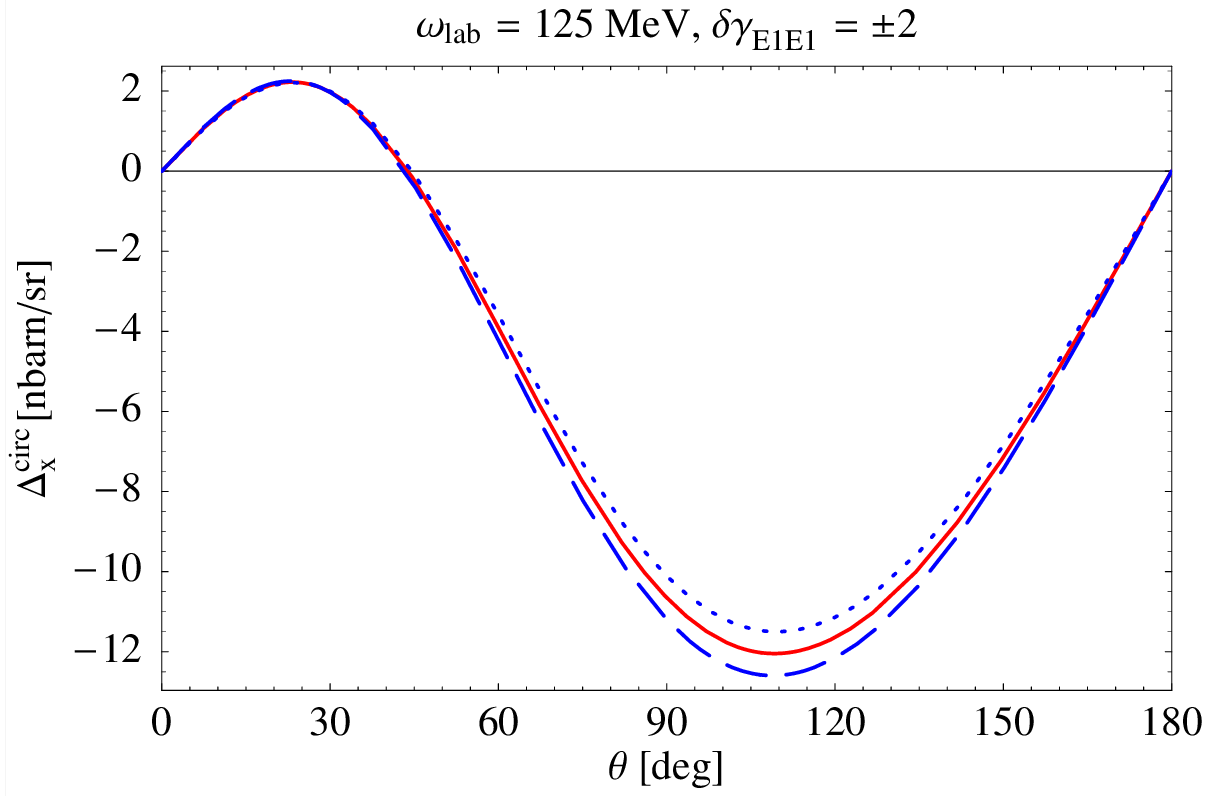}\\[1ex]
\includegraphics*[width=0.31\linewidth]{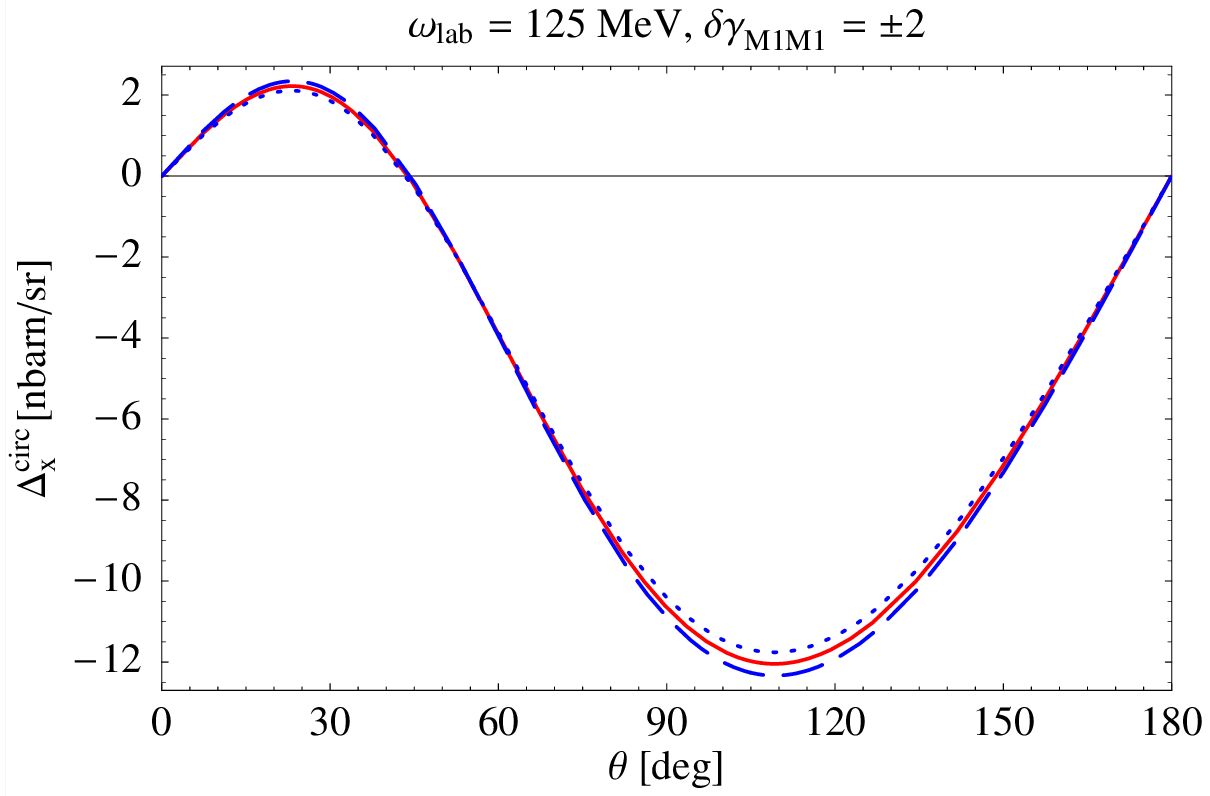}
\hq\hq
\includegraphics*[width=0.31\linewidth]{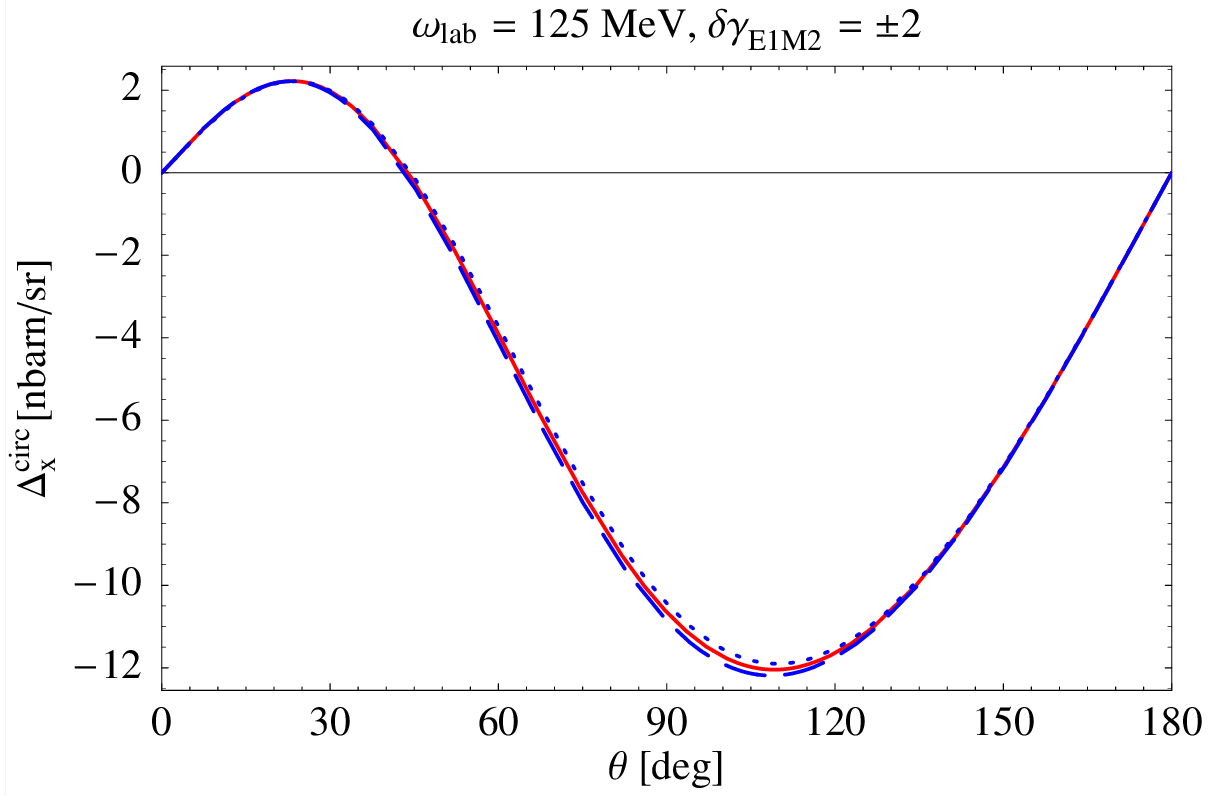}
\hq\hq
\includegraphics*[width=0.31\linewidth]{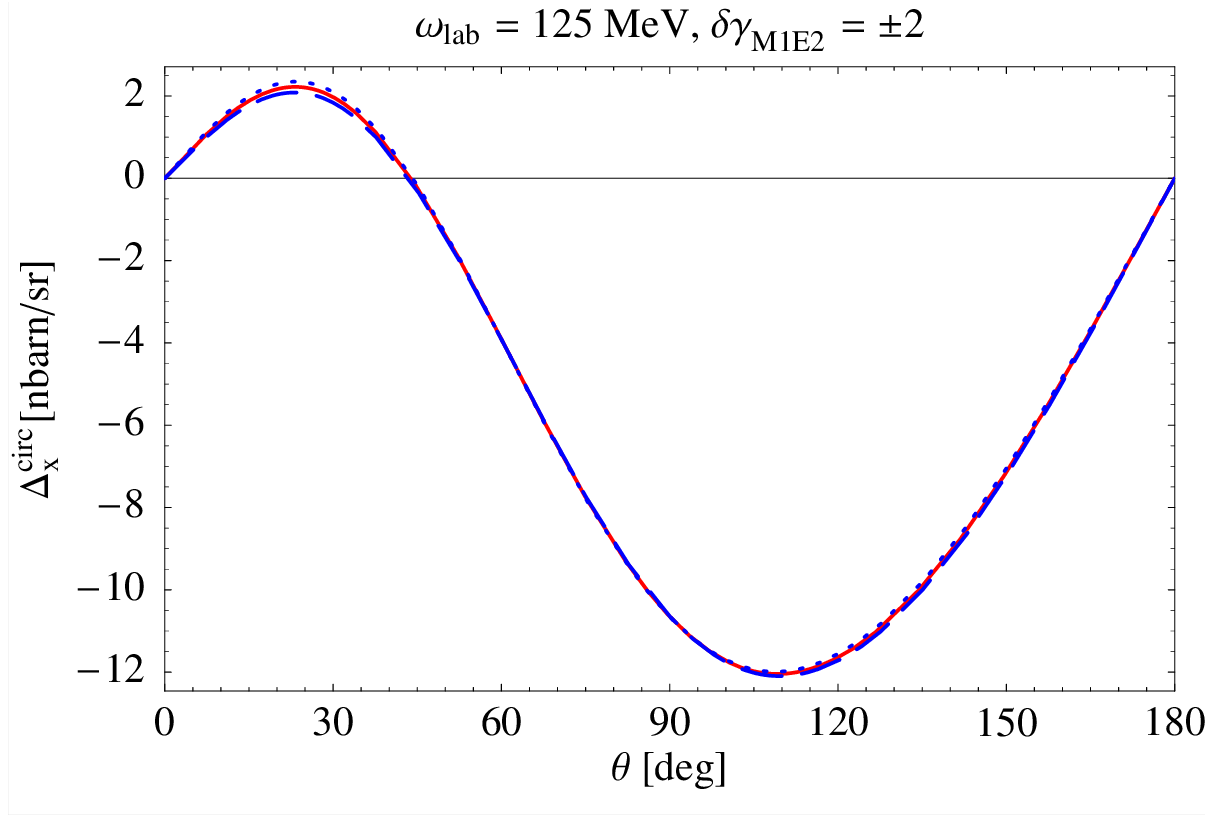}
\caption{\label{fig:deltax} The double-polarization asymmetry
  $\Delta_z^\mathrm{circ}$ with circularly-polarized photons at $\w_\mathrm{lab}=125$
  MeV. From top left to bottom right, variation by $\pm2$ units of
  $\alpha_{E1}$, $\beta_{M1}$, $\gamma_{E1E1}$, $\gamma_{M1M1}$,
  $\gamma_{E1M2}$, $\gamma_{M1E2}$.  }
\end{center}
\end{figure*}

The topmost panel of Fig.~\ref{fig:dcsy} shows $\left.\frac{\dd\sigma}{\dd\W}\right|_y^{lin}$ at 
$45$~MeV (lab). At this energy, there is
appreciable sensitive only to $\alpha_{E1}$ and $\beta_{M1}$.  At 125 MeV (lab) (middle and bottom panels)
however, the sensitivity to $\ga_{M1M1}$ is large and comparable to that of $\al_{E1}-\be_{M1}$. Note that the 
sensitivity to $\ga_{M1M1}$ decreases at back angles where the sensitivity to $\al_{E1}-\be_{M1}$ is largest.
Fig.~\ref{fig:deltazlin} shows the double-polarization observable $\De_z^{lin}$ at $\w_\mathrm{lab}=$ 125~MeV. 
There is comparable sensitivity to three of the polarizabilities$\al_{E1}$, $\be_{M1}$ and $\ga_{M1M1}$.
Finally, we show the double-polarization observable $\De_x^{circ}$ at
$\w_\mathrm{lab}=$ 125~MeV in Fig.~\ref{fig:deltax}. There is appreciable and comparable sensitivity to
$\al_{E1}$, $\be_{M1}$ and $\ga_{E1E1}$, but only minor sensitivity to the
other spin-polarizabilities.

\begin{figure}[!htb]
\begin{center}
\includegraphics*[width=.85\linewidth]{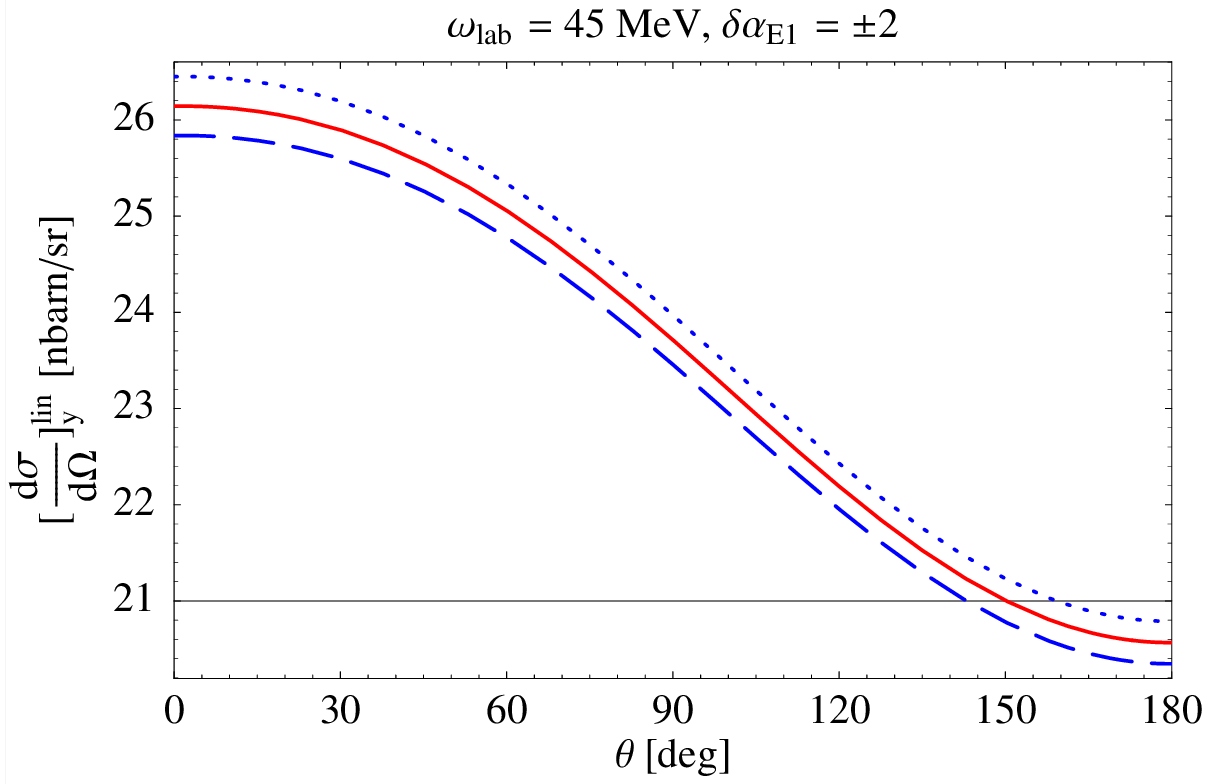}\\
\includegraphics*[width=.85\linewidth]{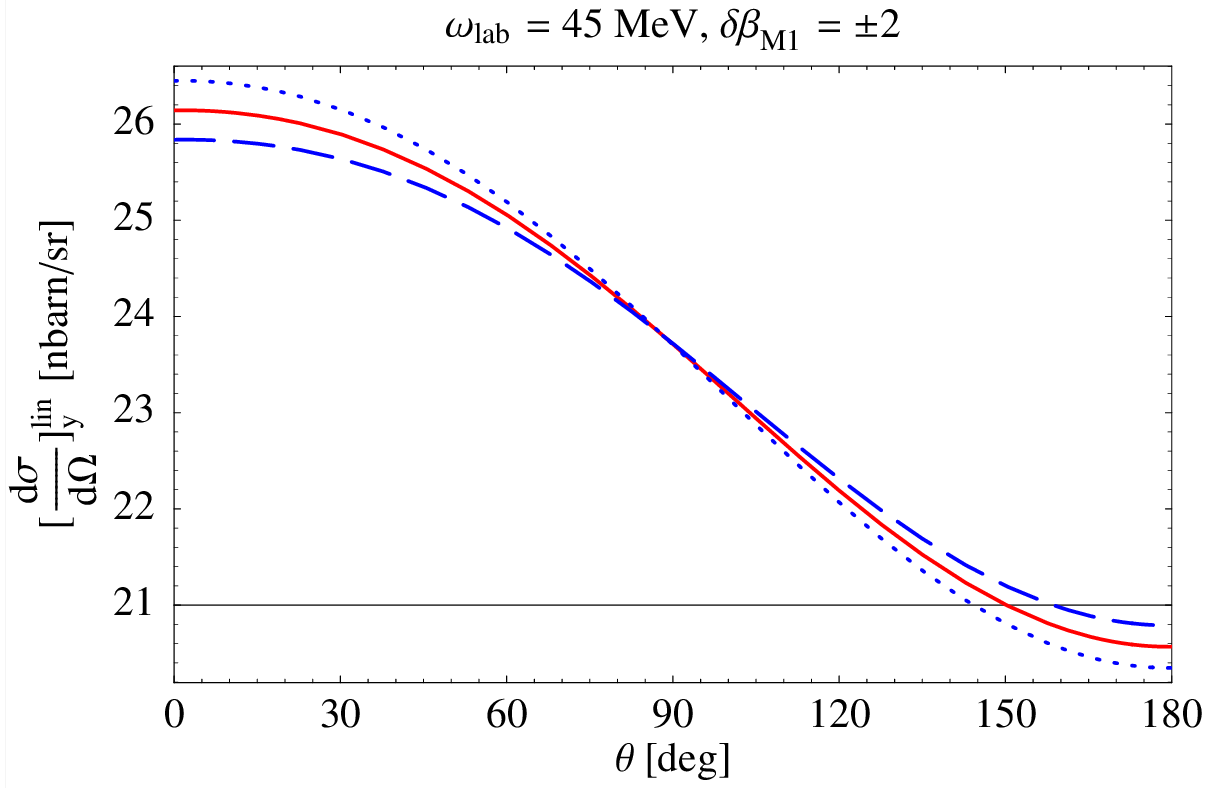}
{\caption {Plots of $\left[ \frac{d\si}{d\W}\right]^{lin}_{y}$ at $\w_{lab}=45$~MeV with varying $\al_{E1}$ 
   (top) and $\be_{M1}$ (bottom).Notice that the sensitivity to $\be_{M1}$ vanishes at 90~deg.}
  \label{fig:dcsy_ab}}
\end{center}
\end{figure}

%\begin{figure*}[!htb]
%\vspace{-2ex}
%\begin{center}
%\parbox{0.13\linewidth}{
%\includegraphics*[width=\linewidth]{configuration2.eps}}%\\
% \mbox{\red{no contribution}~from \blue{$\beta_{M1}$}}}
%\hq\hq\hq\hq
%\parbox{0.64\linewidth}{
%\includegraphics*[width=0.48\linewidth]{dcompton.linpol-sigmay_alpha.38MeV.eps}
%\hq
%\includegraphics*[width=0.48\linewidth]{dcompton.linpol-sigmay_beta.38MeV.eps}}
%\caption{\label{fig:switchingoffpols} Left: Configuration under which a
%  induced dipole cannot radiate an $M1$ photon into the detector.
%  Centre/right: $\left.\frac{\dd\sigma}{\dd\Omega}\right|_y^\mathrm{lin}$ when
%  $\alpha_{E1}$/$\beta_{M1}$ is varied by $\pm2$. Notice the
%  $\beta$-independence at $90^\circ$.}
%\end{center}
%\end{figure*}
The other observables (not shown here) also show minor sensitivity to some of the polarizabilities. The message 
to take away from this analysis is that no obervable is sensitive to only one of the polarizabilities. Ideally,
the extraction of the six polarizabilities should be performed from a global fit to a number of experimental 
measurements. Practically however, one should be judicious in selecting suitable observables for the extraction 
the polarizabilties. Configurations can be isolated where the contribution of one (or more) of the polarizabilities 
is zero. For example, see Fig.~\ref{fig:dcsy_ab} where a detector under $90^\circ$ can therefore not
detect $M1$ photons radiated from the induced magnetic dipole in the nucleon,
cf.~\cite{Maximon:1989zz}. 

Also note that the sensitivity to $\al_{E1}$ and $\be_{M1}$ manifests at much lower energies (see for {\it e.g.}
 Fig.~\ref{fig:dcsy} and \ref{fig:dcsy_ab}) compared to 
the spin polarizabilities. It is therefore advisable to focus on $\al_{E1}$ and $\be_{M1}$ at lower energies 
and the spin polarizabilities at higher energies. It is imperative that the values of the electric and magnetic
polarizabilities be better extracted so as not to taint any extraction of the
spin polarizabilities.

\section{Elastic $\gamma{}^3$He scattering}
\label{sec:gammahe}

This section reports calculations for \he3~Compton scattering at ${\mathcal O}(Q^3)$ in HB\cpt. 
The operator ${\hat O}$ in Eq.~(\ref{calM}) is the irreducible
amplitude for elastic scattering of real photons from the NNN
system. At ${\mathcal O}(Q^3)$
this operator encodes the physics of two photons coupling to a
two-nucleon system inside the \he3~nucleus. We do not have to
include any irreducible three-body Compton mechanisms in our
calculation because they appear at the earliest at ${\mathcal O}(Q^5)$. This allows us to treat the 
\he3~nucleus as a $(2+1)$ nucleon system and enables the simplification of
Eq.~(\ref{calM}) to:
\begin{equation}
{\mathcal M} = 3\bra \Psi_f|\frac{1}{2} \big( {\hat O}^{1B}(1)+{\hat
O}^{1B}(2) \big)+{\hat O}^{2B}(1,2)|\Psi_i\ket , \label{calM2}
\end{equation}
using the Faddeev decomposition of $|\Psi\ket$. The structure of the
calculation is then similar for the one- and two-body parts. The superscript $1B$ in ${\hat O}^{1B}(a)$ 
of Eq. (\ref{calM2}) refers
to the one-body mechanisms or the $\gamma$N amplitude where the
external photon interacts with nucleon `$a$' (refer to Fig.~\ref{fig:oq3N}). Similarly, ${\hat O}^{2B}(a,b)$
represents two-body mechanisms where the external
photons interact with the pair `$(a,b)$' (refer to Fig.~\ref{fig:2bnlographs}).

We calculate ${\mathcal M}$ on a partial-wave Jacobi basis. Convergence of
the results with respect to the angular-momentum expansion was
confirmed. For details on the calculational procedure see
Ref.~\cite{Ch06,Ch07,Sh09}. These \he3~Compton scattering calculations are the first such calculations. 
In future, we intend to extend these calculations to include: the next higher order interactions, the $\De$-isobar
as an explicit degree of freedom and also the effects of intermediate $NNN$-rescattering. 

\subsection{Results}

\begin{figure}[!htb]
\begin{center}
\includegraphics*[width=.85\linewidth]{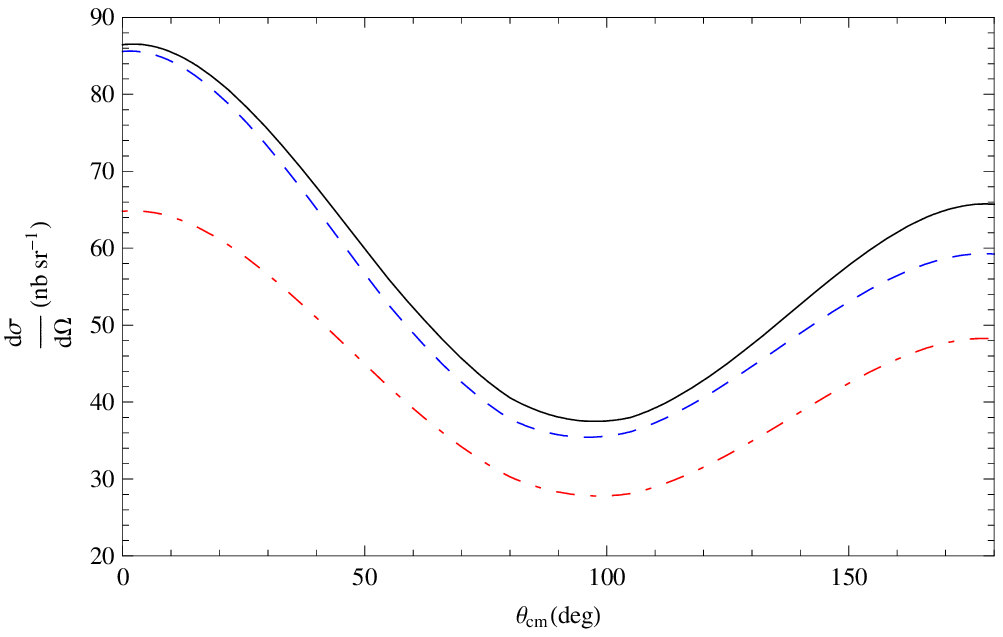}\\
\includegraphics*[width=.85\linewidth]{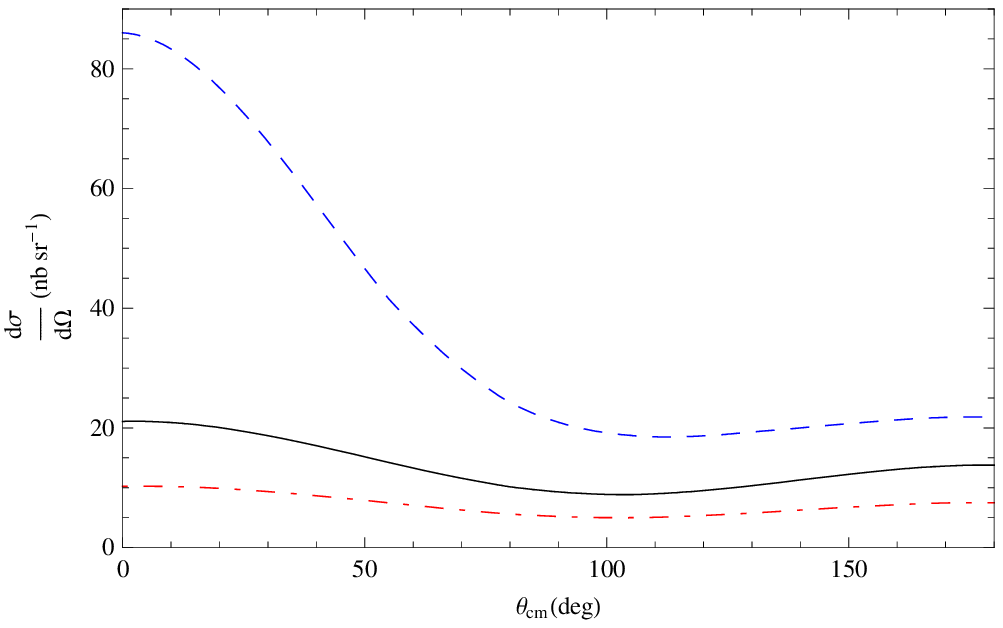}
\caption{\label{fig0} Comparison of different c.m.-frame dcs
calculations at $\w_\mathrm{cm}=$60~MeV (top panel) and $\w_\mathrm{cm}=$120~MeV (bottom panel). The dashed curve is the ${\mathcal O}(Q^2)$ result,
dot-dashed is the IA result and the solid curve is the ${\mathcal O}(Q^3)$ result.}
\end{center}
\end{figure}
%\begin{figure}[htbp]
%\epsfig{figure=061211_prl_dcs_60-120.eps, height=1.4in}
%\caption{\label{fig0} Comparison of different c.m.-frame dcs
%calculations at 60~MeV (left panel) and 120~MeV (right panel).}
%\end{figure}
The amplitude~(\ref{calM2}) is now used to calculate observables. In
Fig.~\ref{fig0} we plot our ${\mathcal O}(Q^3)$ differential cross-section
predictions for elastic $\ga$\he3~scattering. The two panels are
for $\w=$ 60 and 120 MeV. Both show three different dcs
calculations---${\mathcal O}(Q^2)$:dashed curve, $IA$ (Impulse Approximation): dot-dashed curve and
${\mathcal O}(Q^3)$: solid curve. The ${\mathcal O}(Q^2)$ calculation includes
only the proton Thomson term, since that is the LO $\gamma$N amplitude
in $\chi$PT at that order. The $IA$ calculation contains only one-body
${\mathcal O}(Q^3)$ contribution (graphs in Fig.~\ref{fig:oq3N}.
There is a sizeable difference between the
$IA$ and the ${\mathcal O}(Q^3)$ confirming that two-body currents are
equally important. Also, the difference
between ${\mathcal O}(Q^3)$ and ${\mathcal O}(Q^2)$ is very small
at 60 MeV---showing that $\chi$PT may converge well there. This difference 
gradually increases with energy which may be partly because the
fractional effect of the polarizabilities increases with
$\w$.
\begin{figure}[!htb]
\begin{center}
\includegraphics*[width=.85\linewidth]{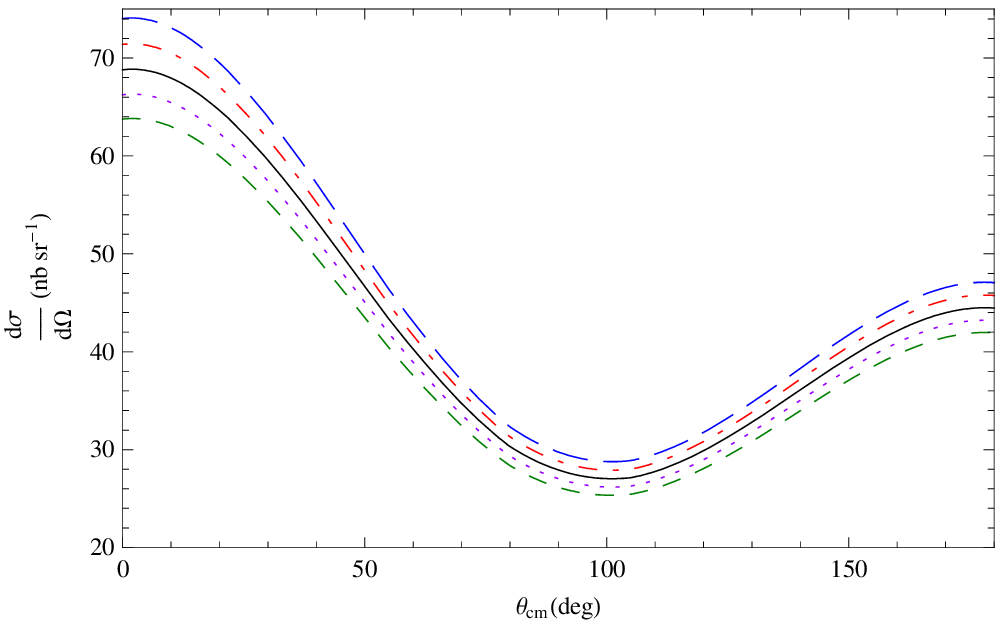}\\
\includegraphics*[width=.85\linewidth]{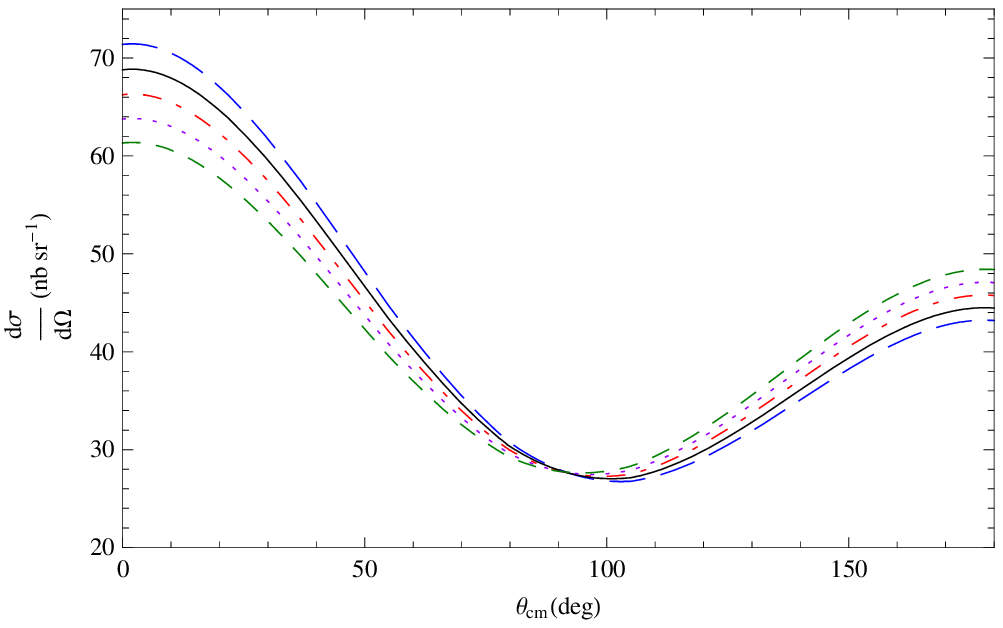}
\caption{\label{fig1} The c.m.-frame ${\mathcal O}(Q^3)$ dcs at $\w_\mathrm{cm}=$80~MeV with varying $\Delta \an$ (left panel) and $\Delta \bn$
(right panel).}
\end{center}
\end{figure}
%\begin{figure}[htbp]
%\epsfig{figure=061221_prl_dcsalphabeta_80.eps, height=1.8in}
%\caption{\label{fig1} The c.m.-frame ${\mathcal O}(Q^3)$ dcs at
%80~MeV with varying $\Delta \an$ (left panel) and $\Delta \bn$
%(right panel).}
%\end{figure}

To analyze the effect of $\al$ and $\be$ on the dcs,  in Fig.~\ref{fig1} we plot the ${\mathcal O}(Q^3)$ dcs at 80~MeV obtained when we add shifts, $\Delta \al^{(n)}$
and $\Delta \beta^{(n)}$, to the ${\mathcal O}(Q^3)$ values of $\al$ and $\be$ for the neutron (see for {\it e.g} 
Ref.~\cite{bkmrev}). $\Delta \an $ is varied in the range $(-4 \ldots 4) \times
10^{-4}$~fm$^3$ (long-dashed curve \ldots dashed curve) in steps of $2 \times
10^{-4}$~fm$^3$. Similarly, $\Delta \bn $ between $(-2 \ldots 6) \times
10^{-4}$~fm$^3$ (long-dashed curve \ldots dashed curve) in steps of $2 \times
10^{-4}$~fm$^3$. This assesses the impact that one set of
higher-order mechanisms has on our ${\mathcal O}(Q^3)$
predictions. Notice that the sensitivity to ${\bn}$ vanishes at $\theta =
90^{\circ}$. This is because $\al^{(n)}$ and $\beta^{(n)}$ enter $A_1^{(n)}$
in the combination $\al^{(n)}+\beta^{(n)}\cos\theta$. This implies that ${\an}$
and ${\bn}$ can be extracted independently from the same experiment.
Secondly, the absolute size of the shift in the dcs due to $\Delta
\an$ and $\Delta \bn$ is roughly the same for all energies. This
suggests that measurements could be done at $\w \approx 80$ MeV,
where the count rate is higher, and the contribution of higher-order
terms in the chiral expansion should be smaller. Thirdly, the sequence of the curves for different values 
of $\Delta \bn $ reverses if one compares the forward and backward angles. This suggests that a meaurement of 
the forward to back angle ratio of the dcs will enhance the effect of $\Delta \bn $.

Before examining double-polarization observables in $\gamma$\he3~
scattering we try to develop some intuition for the $\gamma$\he3~
amplitude. \he3~is a spin-${1\over2}$ target -- this means that the matrix
element~(\ref{calM2}) can be decomposed in the same fashion as 
the nucleon's Compton matrix element in Eq.~(\ref{eq:amp}).
\begin{equation}
T_{\ga ^3He}= \sum \limits_{i=1 \ldots 6} A_i^{^3He}(\w,\theta) t_i;
\;\;\; A_i^{^3He} = A_i^{1B}+A_i^{2B},
 \label{aihe3}
\end{equation}
where $A_i^{1B}$ ($A_i^{2B}$) comes from considering the matrix
element of the one-body (two-body) operators in Eq.~(\ref{calM2}).
The structures $t_3$--$t_6$ now involve the `nuclear' spin. 
In the ground state of polarized \he3, the two proton spins are anti-aligned for the most part,
 therefore the nuclear spin is largely
carried by the unpaired neutron~\cite{he3pol}. We find that the
${\mathcal O}(Q^3)$ two-body currents $A_1^{2B}$ and $A_2^{2B}$
are numerically sizeable, but $A_3^{2B}$--$A_6^{2B}$ are negligible.
Hence, to the extent that polarized \he3~is an `effective neutron', we
expect $A_i^{{}^3He}=A_i^{(n)}$ for $i=3$--$6$. Thus, it is possible to look for the 
effects of the `neutron' spin polarizabilities directly in a polarized \he3~Compton scatteing experiment. 
Moreover, since $A_1^{{}^3He}$ is dominated by the two proton Thomson terms, 
we anticipate a more enhanced signal from
the neutron spin polarizabilities than is predicted for the
corresponding $\ga$d observables (see Sec.~\ref{sec:results} and Refs.~\cite{Ch05,Gr09}).
\begin{figure*}[!htb]
\begin{center}
\includegraphics*[width=.38\linewidth]{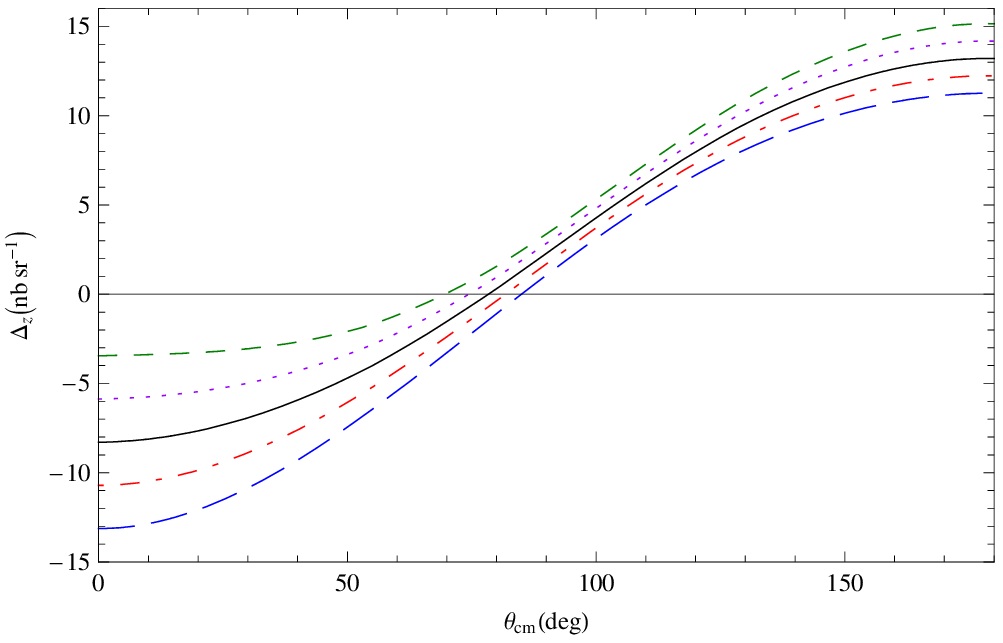}
\includegraphics*[width=.38\linewidth]{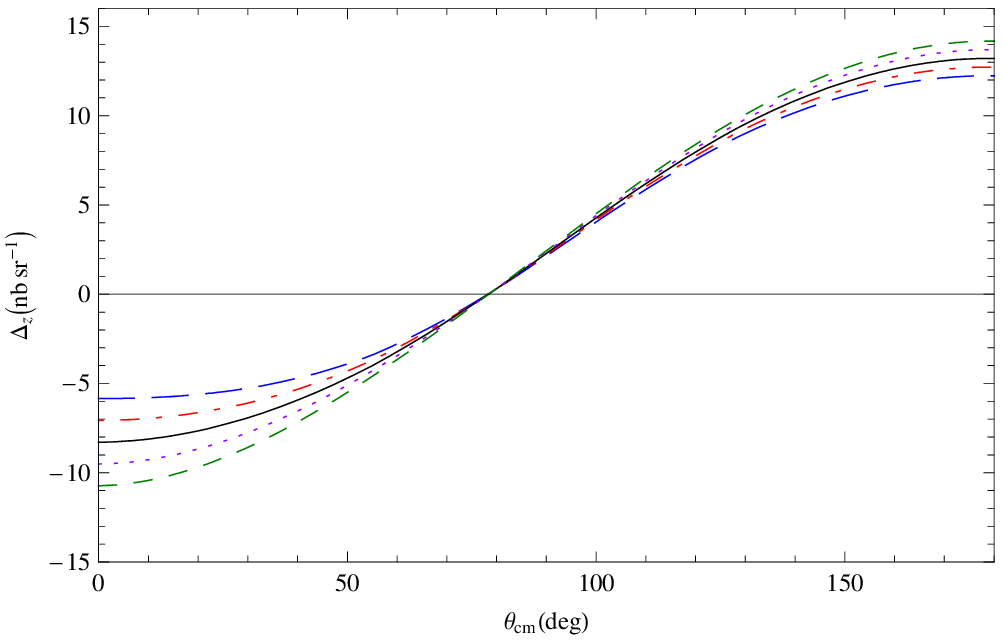} \\
\includegraphics*[width=.38\linewidth]{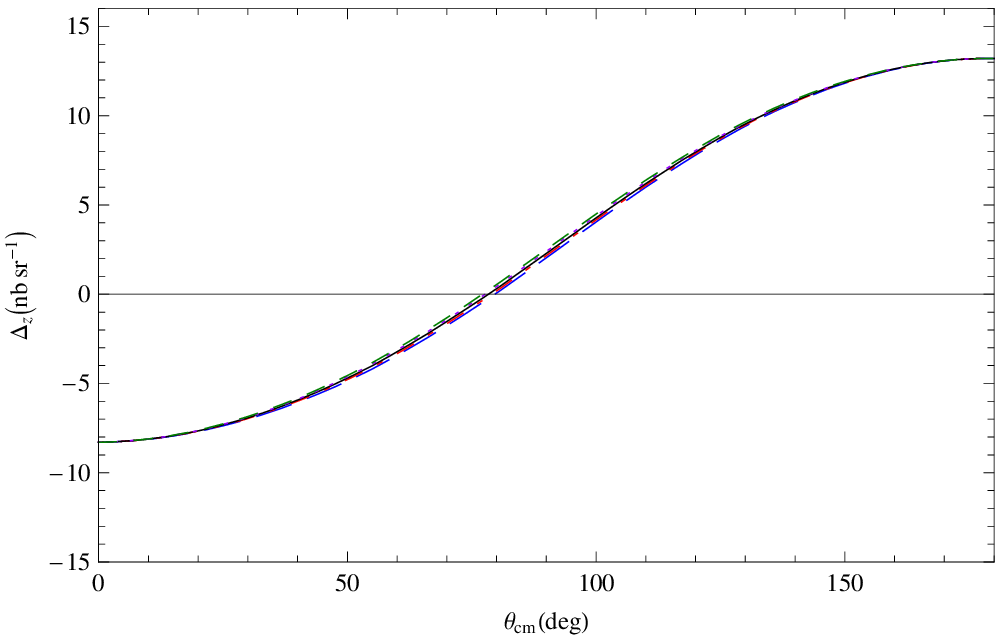}
\includegraphics*[width=.38\linewidth]{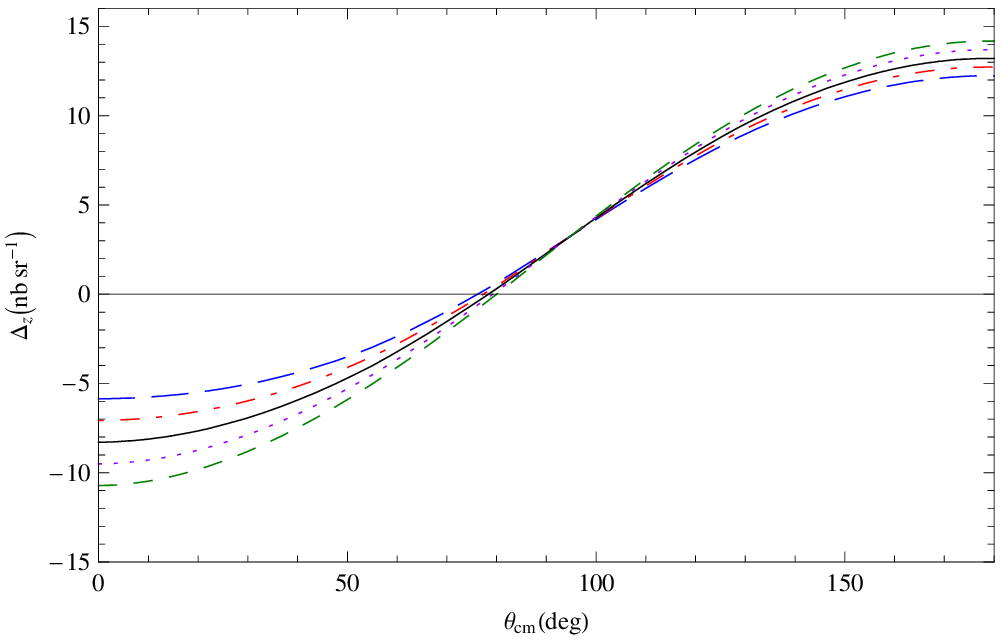}
\caption{\label{fig2} $\Delta_z^\mathrm{circ}$ at $\omega_\mathrm{cm}=$120~MeV with $\ga_1^{(n)}$(top left)--$\ga_4^{(n)}$(bottom right)
varied one at a time. The spin polarizabilities are varied by $\pm 100\%$ of their ${\mathcal O} (Q^3)$
values~\cite{bkmrev}.}
\end{center}
\end{figure*}
%\begin{figure*}[bhtp]
%\epsfig{figure=061204_prl_ddcsztgt_120_gs.eps, height=1.4in}
%\caption{\label{fig2} $\Delta_z^\mathrm{circ}$ at $\omega=$120~MeV with
%(left-to-right) $\ga_1^{(n)}$, $\ga_2^{(n)}$, and $\ga_4^{(n)}$
%varied one at a time. For ${\mathcal O}(Q^3)$ $\ga_i^{(n)}$'s see
%Eq.~(\ref{eq:gammas}).}
%\end{figure*}
\begin{figure*}[!htb]
\begin{center}
\includegraphics*[width=.38\linewidth]{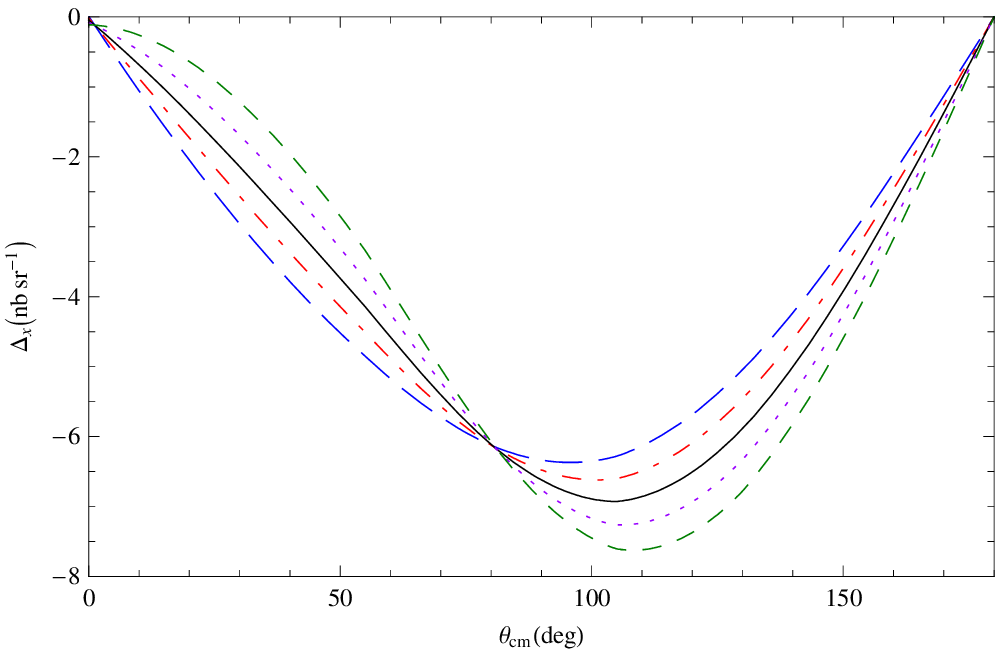}
\includegraphics*[width=.38\linewidth]{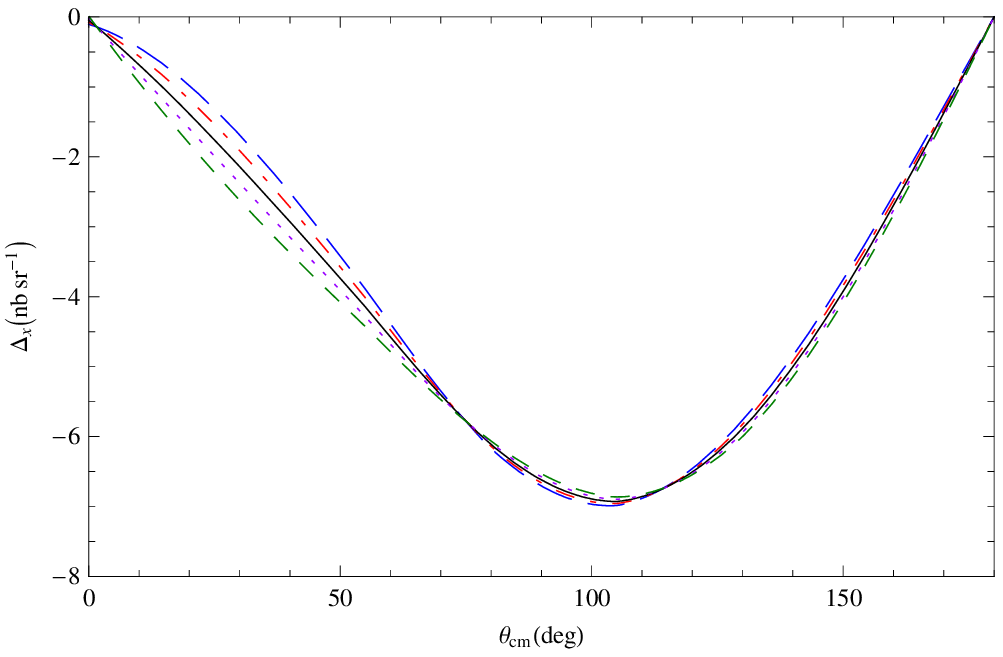}\\
\includegraphics*[width=.38\linewidth]{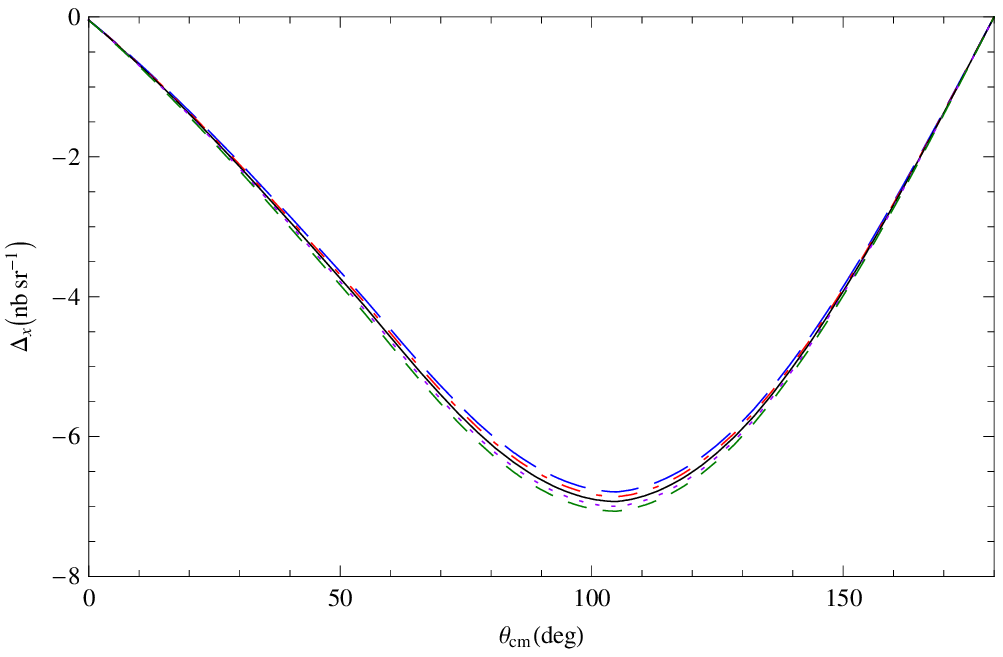}
\includegraphics*[width=.38\linewidth]{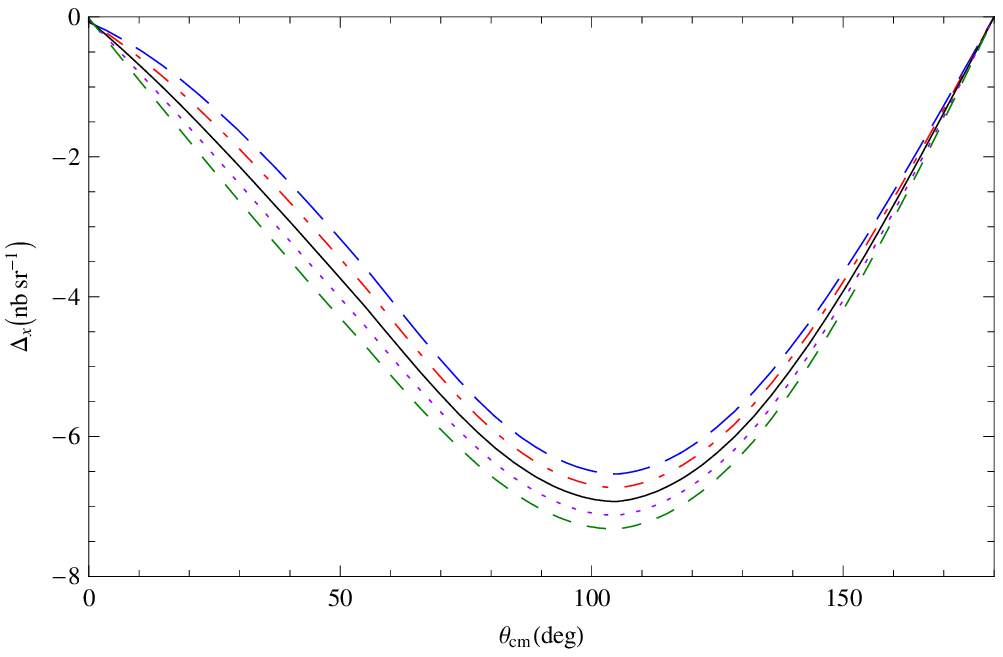}
\caption{\label{fig3} $\Delta_x^\mathrm{circ}$ (c.m. frame) at $\omega_\mathrm{cm}=$120 MeV
when $\ga_1^{(n)}$ (top left)--$ \ga_4^{(n)}$ (bottom right) are varied one
at a time. The spin polarizabilities are varied by $\pm 100\%$ of their ${\mathcal O} (Q^3)$
values~\cite{bkmrev}.}
\end{center}
\end{figure*}
%\begin{figure}[htbp]
%\epsfig{figure=061205_prl_ddcsxtgt_120_gs.eps, height=1.4in}
%\caption{\label{fig3} $\Delta_x^\mathrm{circ}$ (c.m. frame) at $\omega=$120 MeV
%when $\ga_1^{(n)}$ (left) and $ \ga_4^{(n)}$ (right) are varied one
%at a time. Legend as in Fig.~\ref{fig2}.}
%\end{figure}

We emphasize that these arguments are meant only as a guide to the
physics of our exact ${\mathcal O}(Q^3)$ calculation. Our \he3~
wave function is obtained by solving the Faddeev equations with NN
and 3N potentials derived from $\chi$PT. All of the effects due to
neutron depolarization and the spin-dependent pieces of
$\hat{O}^{2B}$ are included in our calculation of the amplitude
(\ref{calM}). This yields the results for $\Delta_z^\mathrm{circ}$ and $\Delta_x^\mathrm{circ}$
shown in Figs.~\ref{fig2} and \ref{fig3}. In these figures we have chosen to vary $\ga_1^{(n)}$, $\ga_2^{(n)}$, and
$\ga_2^{(n)}$ and $\ga_4^{(n)}$. (These are parameterizations of the spin polarizabilities following 
Ragusa~\cite{barry}. $\ga_1^{(n)}$, $\ga_2^{(n)}$, and
$\ga_2^{(n)}$ and $\ga_4^{(n)}$ are linear combinations of the multipole parameterizations of Eq.~(\ref{eq:H2}).)
In both Fig.~\ref{fig2} and Fig.~\ref{fig3} the $\Delta \ga_1^{(n)} \ldots \Delta \ga_4^{(n)}$ (top left to bottom right) 
are varied one by one by $\pm 100\%$ of their $\mathcal{O}(Q^3)$ 
predicted values~\cite{bkmrev}. Both $\Delta_z^\mathrm{circ}$ and $\Delta_x^\mathrm{circ}$
are quite sensitive to $\ga_1^{(n)}$, $\ga_2^{(n)}$, and
$\ga_4^{(n)}$ -- $\Delta_z^\mathrm{circ}$ seems to be sensitive to the combination 
$\ga_1^{(n)} - (\ga_2^{(n)} + 2\ga_4^{(n)})\cos \theta$ whereas $\Delta_x^\mathrm{circ}$ is
sensitive to a different one. With the expected photon flux at an upgraded
HI$\vec{\gamma}$S such effects can be measured~\cite{gao}. Notice that two different 
linear combinations of $\ga_1^{(n)}$, $\ga_2^{(n)}$, and
$\ga_4^{(n)}$ can be extracted through measurements at different angles. For a more detailed
discussion see~\cite{Ch06,Sh09}.  Thus, $\Delta_z^\mathrm{circ}$ and $\Delta_x^\mathrm{circ}$ are
sensitive to two different linear combinations of $\ga_1^{(n)}$,
$\ga_2^{(n)}$, and $\ga_4^{(n)}$ and their measurement should
provide an unambiguous extraction of $\gamma_1^{(n)}$, as well as
constraints on $\gamma_2^{(n)}$ and $\gamma_4^{(n)}$.

\section{Summary and Outlook}
\label{sec:summ}

Results of elastic deuteron and \he3~Compton scattering have been presented with the aim of extracting neutron 
polarizabilities. Our results show that most of the observables are quite sensitive to the polarizabilities. 
The electric and magnetic polarizabilities can also be directly extracted from the unpolarized
deuteron Compton scattering dcs (as shown in Refs.~\cite{Hi05b,Hi05,Be02}) and unpolarized dcs measurements are underway at
MAXLab in Sweden~\cite{myers}. It is essential to accurately know the values of the electric and magnetic
polarizabilities before focussing on the
spin polarizabilities which are `higher-order' effects. The spin polarizabilities start playing 
a role at energies of around 80--90~MeV. Our results have shown that some of the double-polarization 
observables are sensitive to different linear combinations of the spin polarizabilities. 

In view of our results we would like to advocate that--
\begin{enumerate}
 \item Since accurate extraction of the electric and magnetic polarizabilities is central to the plan 
of extracting the six polarizabilities, a number of relatively low-energy experiments ($\lesssim 80$~MeV) 
should focus on extracting $\al^{(n)}$ and $\be^{(n)}$. Apart from the aforementioned
 unpolarized deuteron Compton scattering experiment at MAXLab, an experiment at 
\higs~has also been approved. Planning is also underway for a \he3~Compton scattering experiment at \higs~\cite{gao}.
Using two different nuclei for these measurements would help us better understand the nuclear effects.
Also note that some of the single/double-polarization observables can also 
be used for these extractions. We emphasise that should such measurements be necessary, 
they should be done at lower 
energies where the effects of the spin polarizabilities are negligible.

\item Once $\al$ and $\be$ are better known, a series of simultaneous polarized measurements at 
higher energies ($\gtrsim$ 100 MeV but below the pion-production threshold) would be required to determine the spin 
polarizabilities. The best observables to hunt for the spin polarizabilities seem to be the 
double-polarization observables for both deuteron and \he3~Compton scattering. Measurement of several linear 
combinations of these from a series of experiments can make it ultimately possible to decouple the four 
independent spin polarizabilities.
\end{enumerate}

Ideally, a global fit to a sizeable database that includes polarized/unpolarized measurements from different 
targets, from low to high (but below the pion-production threshold) energies at various angles would result in an 
unambiguous extraction of the polarizabilities. At the other extreme, one can argue that six measurements are 
sufficient to determine the six polarizabilities. However, realistically speaking we need a judicious combination of 
different kinds of measurements. Other experimental avenues may include quasi-free deuteron and He-3 Compton 
measurements.

In the meantime, theoretical efforts should include efforts to improve the accuracy of current calculations. 
As far as deuteron Compton scattering calculations are concerned effort is ongoing to perform a full NNLO
calculation~\cite{allofus} with nucleons, pions and the Delta as
effective degrees of freedom. These would then be the
state-of-the-art systematic calculations describing the deuteron
Compton scattering process from the Thomson limit to the pion
production threshold. The next steps for \he3~Compton scattering would involve (in order of importance)--
\begin{itemize}
\item incorporating the $\De$-isobar as an explicit degree of freedom. As in deuteron Compton scatering we expect the $\De$ to play a 
significant role even at energies $\sim$100~MeV. 
\item Following this, effort needs to be directed towards restoration 
of the Thomson limit in \he3~Compton scattering. This would include resumming the intermediat $NNN$-rescattering states.
\item A complete NNLO extension with nucleons, pions and the $\De$ as explicit degrees of freedom.
\end{itemize}

Lastly, one should remember that these calculations/measurements assume that the proton polarizabilities are 
accurately known. This is true for $\al^{(p)}$ and $\be^{(p)}$, but the poton spin polarizabilities are 
not well-known. Thus, there must be efforts to extract these quantities before concentrating on the 
neutron spin polarizabilities. It is indeed encouraging that polarized proton Compton scattering~\cite{gap}
 measurements are in the pipeline precisely for this reason.

\section{Acknowledgements}
I would like to thank my collaborators H.~Grie{\ss}hammer, J.~McGovern, A.~Nogga and D.~Phillips. This work was supported by the National Science Foundation (CAREER grant PHY-0645498) and
US-DOE (DE-FG02-97ER41019 and DE-FG02-95ER-40907). Furthermore, I would like to thank
the organizers for the generous financial support and an excellent conference.


\begin{thebibliography}{99}

\bibitem{barry}
S.~Ragusa, Phys. Rev. {\bf D47}, 3757 (1993); B.~R. Holstein,
D.~Drechsel, B.~Pasquini, M.~Vanderhaeghen, Phys. Rev. {\bf C61},
034316 (2000).

\bibitem{bkmrev}
V.~Bernard, N.~Kaiser, and Ulf-G.~Mei{\ss}ner, Int. J. Mod. Phys.
{\bf E4}, 193 (1995).

\bibitem{Schumacher06}
M.~Schumacher, Prog. Part. Nucl. Phys. {\bf 55}, 567 (2005).

\bibitem{Hi05b} R.~P. Hildebrandt, H.~W. Grie{\ss}hammer, T.~R. Hemmert,
  {\tt nucl-th/0512063} (2005).

\bibitem{Hi05} R.~P. Hildebrandt, Ph. D. Thesis, \texttt{arXiv.org:nucl-th/0512064}.

\bibitem{Ko03}
 K.~Kossert, {\em et~al.}, Eur. Phys. J., {\bf A16}, 259 (2003).

\bibitem{Pa03}
V. Pascalutsa, and D.~R. Phillips, Phys. Rev. {\bf C68}, 055205
(2003).

\bibitem{ggt}
M.~Gell-Mann, M.~L. Goldberger and W.~E. Thirring, Phys. Rev. {\bf
95}, 1612 (1954).

\bibitem{Sa94}
A.~M. Sandorfi, M. Khandaker, and C.~S. Whisnant, Phys. Rev. {\bf
D50}, R6681 (1994).

\bibitem{myers} 
G.~Feldman, {\em et~al.}, Few Body Sys. {\bf 44}, 325 (2008); L.~Myers, Private communication.

\bibitem{he3pol}
B.~Blankleider and R.~M. Woloshyn, Phys. Rev., {\bf C29}, 538
(1984); J.~L. Friar {\it et al.}, Phys. Rev., {\bf C42}, 2310 (1990).

\bibitem{Ch06}
D.~Choudhury, Ph.D. Thesis, Ohio University, 2006.

\bibitem{Ch07} D.~Choudhury, A.~Nogga and D.~R. Phillips, Phys. Rev. Lett.
  {\bf 98} 232303 (2007).

\bibitem{Sh09}
 D.~Shukla, A.~Nogga and D.~R. Phillips, Nucl. Phys.
  {\bf A819} 98 (2009).

\bibitem{allofus} H.~Grie{\ss}hammer, J.~McGovern, D.~R.~Phillips, D.~Shukla,
  work in progress.

\bibitem{Be02} S.~R. Beane, M.~Malheiro, J.~A. McGovern, D.~R. Phillips,
  U.~van Kolck, Phys. Lett {\bf B567}, 200 (2003). Erratum-ibid: Phys.  Lett.
  {\bf B607} 320 (2005); S.~R. Beane, M.~Malheiro, J.~A. McGovern, D.~R.
  Phillips,  U.~van Kolck, Nucl. Phys. {\bf A747}, 311--361 (2005).

\bibitem{Hi04} R.~P. Hildebrandt, H.~W. Grie{\ss}hammer, T.~R. Hemmert,
  B.~Pasquini, Eur.Phys.J. {\bf A20} 293(2004).

\bibitem{Be99} S.~R. Beane, M.~Malheiro, D.~R. Phillips,  U.~van Kolck,
  Nucl.  Phys. {\bf A656}, 367--399 (1999).

\bibitem{Hi05a} R.~P. Hildebrandt, H.~W. Grie{\ss}hammer, T.~R. Hemmert,
  D.~R.  Phillips, Nucl. Phys., {\bf A748}, 573 (2005).

\bibitem{Ch05}
 D.~Choudhury, and  D.~R. Phillips, Phys. Rev., {\bf C71}, 044002
(2005).

\bibitem{Mc00}
J.~A. McGovern, Phys. Rev. {\bf C63}, 064608 (2001).

\bibitem{He97} T.~R. Hemmert, B.~R. Holstein and J.~Kambor, Phys. Rev.{\bf
    D55}, 5598 (1997); T.~R. Hemmert, B.~R. Holstein, J.~Kambor and
  G.~{Kn\"ochlein}, Phys. Rev. {\bf D57}(9), 5746 (1998).

\bibitem{Friar} J.L.~Friar, Ann.~of Phys.~\textbf{95}, 170 (1975);
  H.~Arenh\"ovel, Z.~Phys.~A\textbf{297}, 129 (1980); M.~Weyrauch and
  H.~Arenh\"ovel, Nucl.~Phys.~A\textbf{408}, 425 (1983).

\bibitem{Maximon:1989zz}
  L.~C.~Maximon,
  Phys.\ Rev.\  C {\bf 39} (1989) 347.

\bibitem{Gr09} H.~Grie{\ss}hammer and D.~Shukla, \texttt{arXiv:nucl-th/0910.0053}; 
D.~Shukla and H.~Grie{\ss}hammer, forthcoming.

\bibitem{gao}
H.~Gao, Private Communication.

\bibitem{gap}
R.~Miskimen, in Proceedings of the 5th International Workshop on
Chiral Dynamics, Theory and Experiment, Durham, NC, 2006, published by World
Scientific, Singapore; H.~Weller \textit{et al.}, Prog. Part. Nucl. Phys., {\bf 62}, 257 (2009).

\end{thebibliography}
\end{document}